\PassOptionsToPackage{prologue,table}{xcolor}
\documentclass[acmtog]{acmart}

\gdef\cropInsets{0}

\usepackage{animate}

\usepackage{xargs}
\usepackage[prologue,table]{xcolor}
\usepackage{booktabs} %
\usepackage{tabularx}
\usepackage{caption}
\usepackage{nicematrix}
\usepackage{float}
\usepackage{mathtools}
\usepackage{amsmath,amsfonts}
\usepackage{bm}
\usepackage{microtype}
\usepackage{units}
\usepackage{subcaption}
\usepackage{overpic}
\usepackage{cancel}
\usepackage{wrapfig}
\usepackage{placeins}
\usepackage{empheq}
\usepackage[export]{adjustbox}
\usepackage[vlined,boxed,ruled]{algorithm2e} %

\SetKw{Or}{or}
\SetKw{And}{and}
\SetStartEndCondition{ }{}{}%
\SetKwProg{Fn}{def}{\string:}{}
\SetKwFunction{Range}{range}%
\SetKw{KwTo}{in}
\SetKwFor{For}{for}{\string:}{}%
\SetKwFor{While}{while}{:}{fintq}%
\SetKwIF{If}{ElseIf}{Else}{if}{:}{elif}{else:}{}%
\SetKwComment{Comment}{\# }{}
\SetKw{Continue}{continue}
\AlgoDontDisplayBlockMarkers
\SetAlgoNoEnd
\SetAlgoNoLine
\DontPrintSemicolon

\SetAlFnt{\small}
\SetAlCapFnt{\small}
\SetAlCapNameFnt{\small}
\SetAlCapHSkip{0pt}
\IncMargin{-\parindent}

\usepackage{xr}
\usepackage{multirow}
\usepackage[capitalize,noabbrev]{cleveref} %
\usepackage{url} %
\usepackage[utf8]{inputenc}

\def\commentType{0}

\ifnum\commentType=0
    \newcommandx{\customComment}[3]{}
    \newcommandx{\customTODO}[3]{}
\fi
\ifnum\commentType=1
    \usepackage[
    type=upperleft,
    unit=in
    ]{fgruler}
    \newcommandx{\customComment}[3]{\textcolor{#2}{\textsl{#1: #3}}}
    \newcommandx{\customTODO}[3]{\textcolor{#2}{\textsl{#1: #3}}}
\fi
\ifnum\commentType=2
    \usepackage[
    type=upperleft,
    unit=in
    ]{fgruler}
    \usepackage{pdfcomment}
    \newcommandx{\customComment}[3]{\pdfcomment[icon=Comment,opacity=0.5,color=#2,author=#1]{#3}}
    \newcommandx{\customTODO}[3]{\pdfcomment[icon=Note,opacity=0.5,color=#2,author=#1]{#3}}
\fi
\ifnum\commentType=3
    \usepackage[
    type=upperleft,
    unit=in
    ]{fgruler}
    \usepackage{pdfcomment} %
    \usepackage{todonotes} %
    \usepackage{silence}
    \newcommandx{\customComment}[3]{\todo[color=#2!40,size=\small]{\textbf{#1:} #3}}
    \newcommandx{\customTODO}[3]{\todo[color=#2!40,size=\small]{\textbf{#1:} #3}}
\fi

\let\originalleft\left %
\let\originalright\right %
\renewcommand{\left}{\mathopen{}\mathclose\bgroup\originalleft} %
\renewcommand{\right}{\aftergroup\egroup\originalright} %

\definecolor{amber}{rgb}{1.0, 0.49, 0.0}
\definecolor{darkgreen}{rgb}{0.0, 0.5, 0.0}
\definecolor{darkblue}{rgb}{0.0, 0.0, 0.5}
\newcommandx{\All}[1]{\customComment{All}{red}{#1}}
\newcommandx{\Ana}[1]{\customComment{Ana}{amber}{#1}}
\newcommandx{\Justin}[1]{\customComment{Justin}{darkgreen}{#1}}
\newcommandx{\justin}[1]{\customComment{Justin}{darkgreen}{#1}}
\newcommandx{\vincent}[1]{\customComment{Vincent}{darkblue}{#1}}
\newcommandx{\Oded}[1]{\customComment{Oded}{darkgreen}{#1}}
\newcommandx{\Miri}[1]{\customComment{Miri}{darkgreen}{#1}}

\newcommandx{\TODO}[1]{\customTODO{TODO}{red}{#1}}
\newcommandx{\AnaTODO}[1]{\customTODO{Ana TODO}{amber}{#1}}
\newcommandx{\JustinTODO}[1]{\customTODO{Justin TODO}{darkgreen}{#1}}
\newcommandx{\OdedTODO}[1]{\customTODO{Oded TODO}{darkgreen}{#1}}
\newcommandx{\MiriTODO}[1]{\customTODO{Miri TODO}{darkgreen}{#1}}

\newcommand{\REMOVE}[1]{} %

\def\equationautorefname~#1\null{%
  Equation~(#1)\null
}

\newcommand{\eqtag}[1]{{\text{{\small{(#1)}}}}}

\newcommand{\suchThat}[0]{{\mathrm{s.t.}\quad}}

\newcommand{\pos}{\bm{x}}

\newcommand{\posBoundary}{\bm{v}}

\newcommand{\DomainSample}[0]{\pos}
\newcommand{\BoundarySample}[0]{\posBoundary}

\newcommand{\bary}[0]{\alpha}
\newcommand{\baryv}[0]{\pmb{\alpha}}

\newcommand{\Domain}{\mathcal{P}}

\newcommand{\Boundary}{\mathcal{V}}
\newcommand{\Triangle}{\mathcal{T}}%

\newcommand{\R}[0]{{\mathbb{R}}}

\newcommand{\TriSet}{\mathbb{T}}

\newcommand{\Diff}[1]{\,\mathrm{d}#1}

\newcommand{\Laplacian}[0]{\Delta}
\DeclareMathOperator{\proj}{proj}
\DeclareMathOperator{\ARAP}{ARAP}
\DeclareMathOperator{\Dir}{Dir}

\newcommand{\bigO}{\mathcal{O}}

\newcommand{\Armadillo}{\textsc{Armadillo}}
\newcommand{\Sappho}{\textsc{Bust of Sappho}}
\newcommand{\Horse}{\textsc{Horse}}

\newcommand{\Gecko}{\textsc{Gecko}}
\newcommand{\Panda}{\textsc{Panda}}
\newcommand{\Human}{\textsc{Human}}
\newcommand{\Elephant}{\textsc{Elephant}}
\newcommand{\Hand}{\textsc{Hand}}
\newcommand{\Star}{\textsc{Star}}
\newcommand{\Woody}{\textsc{Woody}}
\newcommand{\BlueMonster}{\textsc{Blue Monster}}

\gdef\useCroppedImages{1}

\usepackage[export]{adjustbox}
\usepackage{calc}
\usepackage{tabularx}
\usepackage{tikz}
\usepackage{xstring}

\newlength{\beautyHeight}
\newlength{\beautyPixWidth}
\newlength{\beautyPixHeight}
\newlength{\insetvsep}
\gdef\useInsetA{0}
\gdef\useInsetB{0}
\gdef\useInsetC{0}

\newcommand{\setInset}[6]{%
    \expandafter\gdef\csname useInset#1\endcsname{1}%
    \expandafter\gdef\csname inset#1Color\endcsname{#2}%
    \expandafter\gdef\csname crop#1X\endcsname{#3}%
    \expandafter\gdef\csname crop#1Y\endcsname{#4}%
    \expandafter\gdef\csname crop#1W\endcsname{#5}%
    \expandafter\gdef\csname crop#1H\endcsname{#6}%
}

\newcommand{\unsetInset}[1]{%
    \expandafter\gdef\csname useInset#1\endcsname{0}%
}

\newcommand{\addBeautyCrop}[8]{%
    \pdfpxdimen=\dimexpr 1 in/72\relax
    \def\beauty{%
        \let\cropR\relax%
        \let\cropB\relax%
        \newlength\cropR%
        \newlength\cropB%
        \setlength\cropR{{#3 px}-{#5 px}-{#7 px}}%
        \setlength\cropB{{#4 px}-{#6 px}-{#8 px}}%
        \sbox0{\includegraphics[width=#2\textwidth,trim={#5px {\cropB} {\cropR} #6px},clip]{#1}}%
        \begin{tikzpicture}
            \node[anchor=north west,inner sep=0] at (0,0) {\usebox0};
            \begin{scope}[x=\wd0/#7, y=\ht0/#8]
            \if\useInsetA1{
                \draw[\insetAColor,very thick] (\cropAX-#5,-\cropAY+#6) rectangle + (\cropAW,-\cropAH);
            }\fi
            \if\useInsetB1{
                \draw[\insetBColor,very thick] (\cropBX-#5,-\cropBY+#6) rectangle + (\cropBW,-\cropBH);
            }\fi
            \if\useInsetC1{
                \draw[\insetCColor,very thick] (\cropCX-#5,-\cropCY+#6) rectangle + (\cropCW,-\cropCH);
            }\fi
            \end{scope}
        \end{tikzpicture}
    }%
    \setlength\beautyHeight{\heightof{\beauty}}%
    \setlength\beautyPixWidth{#3px}%
    \setlength\beautyPixHeight{#4px}%
    \global\beautyHeight=\beautyHeight%
    \global\beautyPixWidth=\beautyPixWidth%
    \global\beautyPixHeight=\beautyPixHeight%
    \begin{adjustbox}{valign=t}
        \beauty{}
    \end{adjustbox}
}

\newcommand{\trimInset}[6]{%
    \let\cropR\relax%
    \let\cropB\relax%
    \newlength\cropR%
    \newlength\cropB%
    \setlength\cropR{{\beautyPixWidth}-{#3 px}-{#5 px}}%
    \setlength\cropB{{\beautyPixHeight}-{#4 px}-{#6 px}}%
    \color{#2}%
    \fbox{\includegraphics[width=1\linewidth,trim={{#3 px} {\cropB} {\cropR} {#4 px}},clip]{#1}}%
}

\newcommand{\addInset}[2]{%
    \color{#2}%
    \fbox{\includegraphics[width=1\linewidth]{#1}}%
}

\newcommand{\auxtimes}{x}
\newcommand{\auxplus}{+}
\newcommand{\auxspace}{ }

\newcommand{\addInsets}[1]{%
    \begin{adjustbox}{valign=t}
        \StrSubstitute{#1}{.}{-}[\baseFileName]
        \begin{adjustbox}{totalheight=1\beautyHeight,tabular={c}}
            \if\useInsetA1%
                \def\cropfile{\baseFileName-\cropAW\auxtimes\cropAH\auxplus\cropAX\auxplus\cropAY-crop}
                \if\cropInsets1
                    \immediate\write18{convert #1 -crop \cropAW\auxtimes\cropAH\auxplus\cropAX\auxplus\cropAY\auxspace -filter point -resize 800\% \cropfile.png}
                \fi
                \if\useCroppedImages1
                    \addInset{\cropfile.png}{\insetAColor}
                \else
                    \trimInset{#1}{\insetAColor}{\cropAX}{\cropAY}{\cropAW}{\cropAH}%
                \fi%
            \fi%
            \if\useInsetB1%
                \if\useInsetA1\\[\insetvsep]\fi%
                \def\cropfile{\baseFileName-\cropBW\auxtimes\cropBH\auxplus\cropBX\auxplus\cropBY-crop}
                \if\cropInsets1
                    \immediate\write18{convert #1 -crop \cropBW\auxtimes\cropBH\auxplus\cropBX\auxplus\cropBY\auxspace -filter point -resize 800\% \cropfile.png}
                \fi
                \if\useCroppedImages1
                    \addInset{\cropfile.png}{\insetBColor}
                \else
                    \trimInset{#1}{\insetBColor}{\cropBX}{\cropBY}{\cropBW}{\cropBH}%
                \fi%
            \fi%
            \if\useInsetC1%
                \if\useInsetB1\\[\insetvsep]\fi%
                \def\cropfile{\baseFileName-\cropCW\auxtimes\cropCH\auxplus\cropCX\auxplus\cropCY-crop}
                \if\cropInsets1
                    \immediate\write18{convert #1 -crop \cropCW\auxtimes\cropCH\auxplus\cropCX\auxplus\cropCY\auxspace -filter point -resize 800\% \cropfile.png}
                \fi
                \if\useCroppedImages1
                    \addInset{\cropfile.png}{\insetCColor}
                \else
                    \trimInset{#1}{\insetCColor}{\cropCX}{\cropCY}{\cropCW}{\cropCH}%
                \fi%
            \fi%
        \end{adjustbox}
    \end{adjustbox}
}

\definecolor{cartoPrismTeal}{rgb}{0.21960784 0.65098039 0.64705882}
\definecolor{cartoPrismOrange}{rgb}{0.88235294 0.48627451 0.01960784}
\definecolor{cartoPrismGreen}{rgb}{0.45098039 0.68627451 0.28235294}
\definecolor{cartoPrismRed}{rgb}{0.8 0.31372549 0.24313725}
\definecolor{cartoPrismPurple}{rgb}{0.58039216 0.20392157 0.43137255}

\definecolor{mathematicaBlue}{rgb}{0.38, 0.51, 0.71}
\definecolor{mathematicaOrange}{rgb}{0.88, 0.61, 0.14}
\definecolor{mathematicaGreen}{rgb}{0.56, 0.69, 0.19}
\definecolor{mathematicaRed}{rgb}{0.92,0.39, 0.21}
\definecolor{mathematicaPurple}{rgb}{0.53, 0.47, 0.7}

\usepackage{enumitem,enumerate}
\setlist[itemize]{noitemsep, nolistsep, leftmargin=*}

\setcopyright{rightsretained}
\copyrightyear{2023}
\acmYear{2023}
\acmJournal{TOG}
\acmYear{2023} \acmVolume{42} \acmNumber{6} \acmArticle{255} \acmMonth{12} \acmPrice{}\acmDOI{10.1145/3618403}

\acmSubmissionID{papers\_937s2}

\begin{document}

\title[VBC]{Variational Barycentric Coordinates}

\author{Ana Dodik}
\email{anadodik@mit.edu}
\orcid{1234-5678-9012}
\affiliation{%
  \institution{MIT CSAIL}
  \country{USA}
}

\author{Oded Stein}
\orcid{1234-5678-9012}
\affiliation{%
  \institution{University of Southern California and MIT CSAIL}
  \country{USA}
}

\author{Vincent Sitzmann}
\orcid{1234-5678-9012}
\affiliation{%
  \institution{MIT CSAIL}
  \country{USA}
}

\author{Justin Solomon}
\orcid{1234-5678-9012}
\affiliation{%
  \institution{MIT CSAIL}
  \country{USA}
}

\renewcommand{\shortauthors}{Dodik et al.}

\begin{abstract}
We propose a variational technique to optimize for generalized barycentric coordinates that offers additional  control compared to existing models. Prior work represents barycentric coordinates using meshes or closed-form formulae, in practice limiting the choice of objective function. In contrast, we directly parameterize the continuous function that maps any coordinate in a polytope's interior to its barycentric coordinates using a neural field. This formulation is enabled by our theoretical characterization of barycentric coordinates, which allows us to construct neural fields that parameterize the entire function class of valid coordinates. We demonstrate the flexibility of our model using a variety of objective functions, including multiple smoothness and deformation-aware energies; as a side contribution, we also present mathematically-justified means of measuring and minimizing objectives like total variation on discontinuous neural fields. We offer a practical acceleration strategy, present a thorough validation of our algorithm, and demonstrate several applications.
\end{abstract}


\begin{teaserfigure}
  
    \setlength{\fboxrule}{20pt}%
\setlength{\insetvsep}{30pt}%
\renewcommand{\arraystretch}{1}%
\small%
\begin{tabular}{cccc}
    \setInset{A}{cartoPrismRed}{700}{1090}{285}{285}%
    \setInset{B}{cartoPrismPurple}{1167}{511}{276}{276}%
    \setInset{C}{cartoPrismOrange}{772}{689}{343}{343}%
    \addBeautyCrop{figures/teaser/armadillo-tv-composite}{0.36}{1920}{1920}{0}{5.33}{1867.1819}{1741.3726}\hspace{-2.1\tabcolsep-1.5pt} &%
    \addInsets{figures/teaser/armadillo-tv} & %
    \addInsets{figures/teaser/armadillo-arap} &%
    \hspace{-3\tabcolsep}\addBeautyCrop{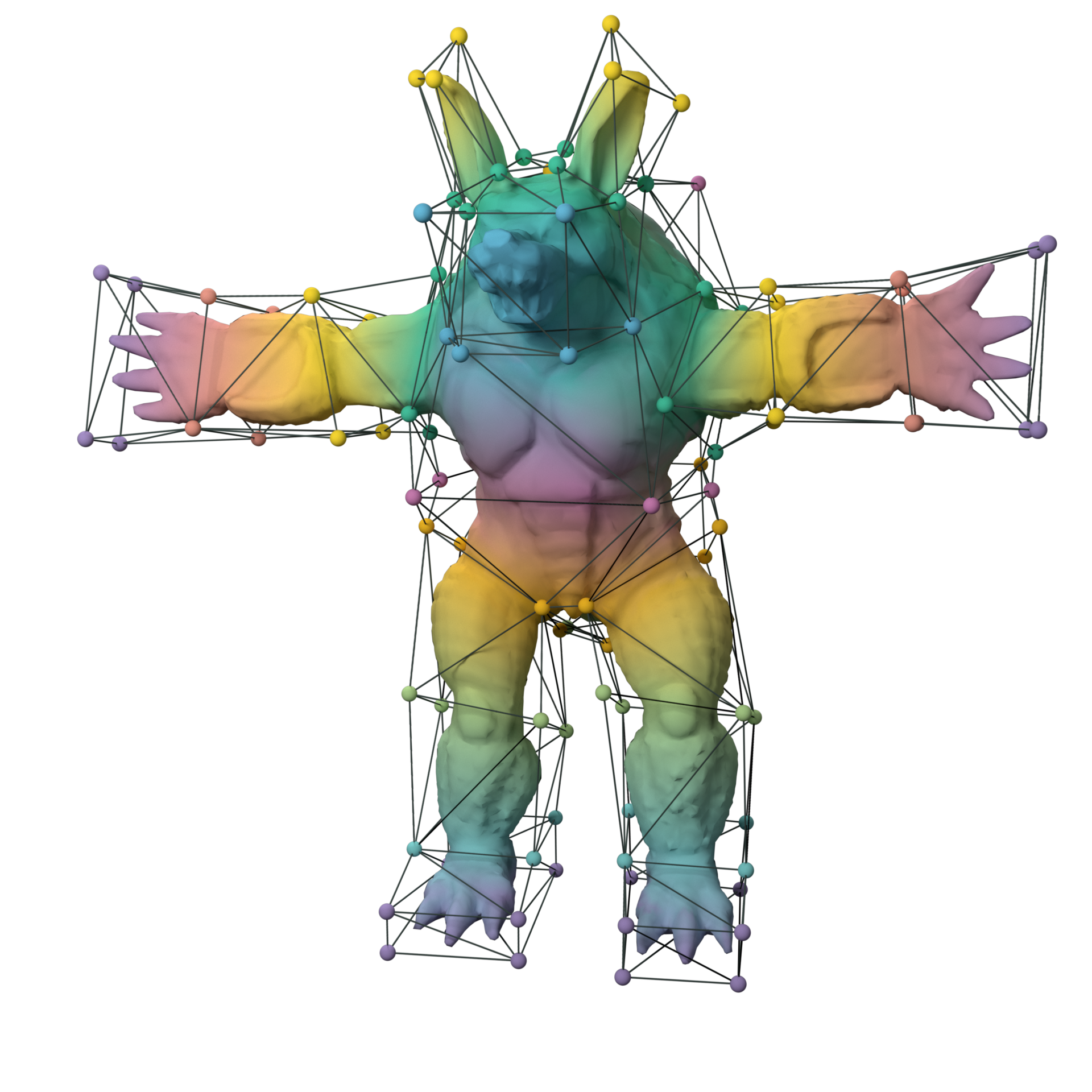}{0.36}{1920}{1920}{0}{5.33}{1867.1819}{1741.3726}\vspace{5pt}\\%
  \cmidrule(lr){1-2} \cmidrule(lr){3-4} \vspace{-10pt}\\%
  \multicolumn{2}{r}{Total Variation} \hspace{4pt} & \multicolumn{2}{l}{\hspace{4pt} Surface ARAP}\vspace{-7pt}%
\end{tabular}%

  \caption{Our method allows us to optimize various energies resulting in different sets of barycentric coordinates. This figure shows the result of using different sets of coordinates on the tasks of color interpolation and mesh deformation. Minimizing \emph{total variation} (TV) results in deformations that have an undesirable rubbery appearance. 
  However, our formulation allows us to address this issue by minimizing \emph{deformation-aware} energies, such as the \emph{as-rigid-as-possible} (ARAP) energy.}
  \Description{This is the teaser figure for the article.}\label{fig:teaser}
\end{teaserfigure}  

\maketitle

\section{Introduction}

Generalized barycentric coordinates are used to interpolate functions defined on the vertices of a polytope to its interior, such as to evaluate a function known only on the surface of a shape inside its volume. They are also often used for \emph{cage-based deformation}, with the polytope serving as a \emph{deformation cage}. In this setting, a user deforms a mesh by moving the vertices of a low-resolution cage that surrounds it. The positions of the transformed cage vertices are then interpolated into its interior, and in particular onto the surface of the mesh inside of it. Figure~\ref{fig:teaser} demonstrates how generalized barycentric coordinates can interpolate data and deform shapes using a cage.

The barycentric coordinates associated to each point in the interior of the cage can be understood as a set of averaging weights, one per cage vertex.
These weights satisfy a number of mathematical constraints: they must be non-negative, sum to one, and have a prescribed expectation.
This final \emph{reproduction constraint} distinguishes barycentric coordinates from more general \emph{skinning weights}; to ensure that a linear deformation of the cage vertices leads to a linear deformation field inside the cage, every point in the interior must equal the linear combination of the cage vertices with its barycentric coordinates as coefficients.
Moreover, we might expect barycentric weights to be smooth and/or local, although these considerations may be understood as objectives rather than hard constraints.

Existing methods for computing barycentric coordinates typically fall into two categories.  Classically, a number of barycentric coordinate functions are expressed in closed-form \cite{Ju2005, Lipman2007} or via relatively simple algorithms.  These coordinates are fast to evaluate but often add assumptions on the cage vertices (e.g., convexity) to satisfy the required properties of barycentric coordinates; moreover, they are inflexible in the sense that they each provide a single means of computing coordinates rather than allowing users to optimize for coordinates best suited for a given application.  
More recently, \citet{Joshi2007} and \citet{Zhang2014} pose the computation of generalized barycentric coordinates as a convex optimization problem.  This approach promises global satisfaction of the constraints defining barycentric coordinate functions and suggests the possibility of optimizing for customized barycentric weights for a given application or artistic intention, but the current models rely on a mesh-based discretization and optimization technique customized to a specific smoothness objective functions.

In this paper, we introduce \emph{variational barycentric coordinates} (VBCs), a flexible mesh-free framework that allows us to optimize for generalized barycentric coordinates given a boundary cage and a differentiable objective. VBCs are built on a simple mathematical observation, namely that all generalized barycentric coordinates can be expressed as weighted averages of simplex barycentric coordinates of all the simplices generated by connecting cage vertices.
With this observation in place, we can represent the set of generalized barycentric coordinates for a given cage using the machinery of neural networks, which allow us to efficiently operate in the high-dimensional space of all possible simplices.
Beyond total variation, we can then optimize using customized objectives tailored to different tasks and applications.

VBCs offer several departures from and advantages over classical generalized barycentric coordinate functions.
Most importantly, our formulation is general and operates in the space of valid barycentric coordinates by construction.
We offer a differentiable representation of the \emph{entire} function class of barycentric coordinate functions, as well as some examples for the possible design choices one could make by incorporating different optimization objectives. %
We also offer a mathematically justified means of computing and optimizing the total variation of our model in the presence of possible discontinuities.

Furthermore, our formulation is defined in terms of the cage vertices, allowing us to support cages composed of meshes, triangle soups, or point clouds. When it is acceptable to restrict to fewer degrees of freedom at the cost of excluding some possible coordinate functions, we offer a practical modification to the algorithm that not only enforces locality but also significantly reduces the memory and compute required.

To demonstrate the benefits of our approach, we provide a thorough qualitative evaluation as well as comparisons with previous work. We additionally offer a quantitative analysis of the different performance trade-offs introduced by our approach. 

\paragraph*{Contributions.} In summary, we introduce:
\begin{itemize}
\item a mathematical formulation that expresses generalized barycentric coordinates as convex combinations of simplex  coordinates,
\item a computational model that uses our formulation to constrain neural networks to the function space of barycentric coordinates,
\item heuristics that help make our model practically tractable while simultaneously enforcing a notion of locality in our coordinates,
\item a way of approximating and optimizing several common smoothness energies---such as total variation and the Dirichlet energy---in the context of discontinuous neural fields,
\item experiments that demonstrate how our coordinates can be combined with familiar \emph{deformation-aware} energies or used to solve inverse deformation problems, and
\item a thorough evaluation and comparison that confirms the validity and practical usefulness of our model on a variety of $2$D and $3$D shapes.
\end{itemize}

\section{Related Work}

\subsection{Generalized Barycentric Coordinates}

Many methods have been proposed to compute generalized barycentric coordinates.  A complete survey is outside the scope of our discussion; we refer the reader to existing surveys \cite{floater-2015, Hormann:2017:GBC} for a comprehensive introduction.  Here, we mention a few particularly relevant works to our formulation.

Barycentric coordinates go back at least as far as M\"obius \cite[p.\ 217]{coxeter1969}, who used them to define a coordinate system on the inside of a triangle, parametrized by the triangle vertex positions.
They have since been generalized to polygons \cite{Floater2003} and higher dimensions \cite{Ju2005,Floater2005}.

In graphics and geometry processing, barycentric coordinates are used for interpolation and deformation.
The cage polygon/polyhedron parameterizes motions of a complicated interior domain in $2$D/$3$D \cite{Lipman2007,Huang2006,Lipman2008,Weber2009,Hormann2008,Li2013,Deng2020,Chen2010}.
Beyond these basic applications, barycentric coordinates are a part of methods for distance computation \cite{Rustamov2009}, image registration \cite{Weistrand2015}, mesh generation \cite{Gregson2011}, finite elements \cite{Wicke2007}, and subdivision \cite{Liu2020}.

Later works on generalized barycentric coordinates use the geometry of the domain's interior to produce coordinates that are aware of local distances; this approach yields more natural-looking animations \cite{Joshi2007}. Unlike early attempts to define generalized barycentric coordinate functions, many of these works pose computation of the coordinates as an optimization problem over the space of possible coordinate functions. 
The recent Local Barycentric Coordinates \cite{Zhang2014} are geometry-aware barycentric coordinates obtained via minimizing an energy containing the Total Variation (TV) of the coordinate functions.
This method was recently accelerated by \citet{Tao2019}.
\citet{Yu2015} optimize the Laplacian energy with a modified boundary term to produce cage-free barycentric coordinates with simple control vertices instead of control cages.
\citet{Stein2018} construct cage-free barycentric coordinates using the Hessian energy with natural boundary conditions, later generalizing the method to curved surfaces \cite{Stein2020}.

In general, closed-form expressions of geometry-aware, locally supported barycentric coordinates are difficult to obtain.
\citet{Anisimov2017} provide closed-form locally-supported barycentric coordinates using a Delaunay triangulation of the cage polygon, but their method does not support polyhedra in 3D.

Lastly, \citet{FLOATER1997} formulates one simple choice of barycentric coordinates in the context of mesh parameterizations as a special case of the formulation presented in \S\ref{ss:model}. Besides the different application domain, Floater's formulation does not support polytopes and is not guaranteed to satisfy the necessary properties on concave polygons (see Figure~\ref{fig:prune} for a failure case). Importantly, our method relies on a neural representation together with an optimization procedure, whereas \citet{FLOATER1997} uses a closed-form formula. This is a key component of our method, as one of our main goals is to enable users to tune weights using arbitrary objectives.

\subsection{Neural Networks for Interpolation and Deformation}

Recent work has demonstrated the potential of treating fully-connected networks as continuous, memory-efficient representations of general functions, shape parts, objects, or scenes by mapping each coordinate to a value stored at that coordinate. These networks are commonly referred to as \emph{neural fields}~\cite{xie2022neural}.
Our method parametrizes barycentric coordinates using neural fields.

In geometry processing, neural fields have been used to parameterize distributions over shape boundary vertices for Linear Blend Skinning (LBS) \cite{jeruzalski2020nilbs}; note that LBS weights are \emph{not} the same as barycentric coordinates, as they do not satisfy the reproduction property. 
Neural fields can also be used to forgo mesh discretizations entirely and perform geometry processing tasks on the field directly \cite{Yang2021}.
\citet{Yifan2020} use a neural network to learn cage-based shape deformations.
Neural representations are also popular to solve physics problems formulated as PDEs on a variety of geometries \cite{Raissi2019,Rao2021,Li2021,Sukumar2022}.
Deforming Neural Radiance Fields (NeRFs) is a popular use-case for cage-based deformation \cite{Peng2021, Yuan2022}, with applications such as modeling dynamic human bodies \cite{Peng2021b} or reconstructions of scenes with deformation \cite{Park2021}.

Beyond neural fields, \citet{Tan2018} use variational autoencoders to model mesh deformation.
\citet{Luo2020} model linear elasticity using neural networks.
\citet{Jiang2020} learn deformations by learning flows between shapes.
\citet{Chentanez2020} model deformations on triangle meshes using convolutional neural networks.
\citet{Aigerman2022} learn intrinsic mappings between meshes using neural networks.

\section{Preliminaries}\label{s:prelim}

We will begin by introducing the broad mathematical definition of generalized barycentric coordinates for an arbitrary polytope cage ${\Domain \subset \R^d}$ with $K$ boundary vertices ${\Boundary(\Domain) = \left\{\BoundarySample_i \mid 1 \leq i \leq K \right\}}$. We also include a brief review of the special case of barycentric coordinates of triangles and tetrahedra in this section, as we will be using them as a building block for our model.

We write \emph{barycentric coordinate functions} for a polytope $\Domain$ as functions ${\bary_i : \Domain \to \R_+}$ and denote by ${\baryv = \left[\bary_1,\dots,\bary_K\right]^\top}$ the corresponding vector-valued barycentric coordinate function.
For $\baryv$ to be useful as interpolation weights, the definition of barycentric coordinates includes a number of constraints that the functions must satisfy~\cite{Floater2006AGC, floater-2015}. In particular, a function $\baryv$ is said to be \emph{valid}
if it fulfills the barycentric coordinate constraints for all $\DomainSample\in\Domain$:
\vspace*{0.1em}
\begin{itemize}
  \item \textsc{Non-negativity.}\; $\bary_i(\DomainSample) \geq 0$;
  \item \textsc{Partition of unity.}\; $\sum_i\bary_i(\DomainSample) = 1$;
  \item \textsc{Reproduction.}\; $\sum_i \BoundarySample_i\bary_i(\DomainSample) = \DomainSample$; and
  \item \textsc{Lagrange property}\; $\bary_i(\BoundarySample_j) = \delta_{ij}$,
\end{itemize}
\vspace*{0.2em}
where \(\delta_{ij}\) is the Kronecker delta.
Previous work has additionally identified \emph{locality}---the idea that coordinates should be non-zero only within a small neighborhood of their vertices---as an optional desirable property for barycentric coordinates \citep{Zhang2014}.

As an example use case for generalized barycentric coordinates, suppose we deform the polytope vertices $\BoundarySample_i$ to new positions $\BoundarySample_i'$.  We can extend this deformation to interior points $\DomainSample$ via the following map:
\begin{equation}\label{eq:deform}
    \varphi_{\baryv}(\DomainSample;\; \Domain') = \sum\nolimits_{i=1}^K \bary_i(\DomainSample) \BoundarySample_i'.
\end{equation}
Thanks to the Lagrange property, we have $\varphi_{\baryv}(\BoundarySample_i)=\BoundarySample_i'$ for all boundary vertices  $i$.  Moreover, thanks to the reproduction property, when $\BoundarySample_i=\BoundarySample_i'$ for all boundary vertices $i$, the map becomes the identity:  $\varphi_{\baryv}(\DomainSample)=\DomainSample.$

\paragraph*{Simplex barycentric coordinates.} For triangles and tetrahedra, this definition leads to unique barycentric coordinates.
In particular, for a triangle in the plane, $\Triangle \subset \R^2$ with vertices ${\Boundary(\Triangle) = \{\BoundarySample_1, \BoundarySample_2, \BoundarySample_3\}}$, and a point $\DomainSample \in \Domain$, the triangle barycentric coordinates are the solution to the linear system
\begin{equation} \label{eq:tribary}
  \begin{bmatrix}
    \BoundarySample_1 & \BoundarySample_2 & \BoundarySample_3 \\
    1 & 1 & 1
  \end{bmatrix}\;
  \baryv(\DomainSample; \Triangle) = 
  \begin{bmatrix}
    \DomainSample \\
    1
  \end{bmatrix}.
\end{equation}
As long as $\DomainSample$ is inside the triangle and the $\BoundarySample_i$ are affinely independent, the solution is unique and satisfies the desired constraints.

In the general case of an arbitrary cage, the linear system above usually is underdetermined, and therefore, the barycentric coordinate functions are not unique.
Instead, they lay somewhere in the feasible set defined by the constraints.
The feasible set $\mathcal{A}$ of generalized barycentric coordinates at a point $\DomainSample$ is a $(K - d - 1)$-dimensional simplex. %
To ensure that our computed coordinates always lie in $\mathcal{A}$, we need to optimize in the constrained space of feasible coordinate functions rather than the larger space of all possible smooth functions.

\begin{figure*}[t!]
  \centering
  \includegraphics[width=\linewidth]{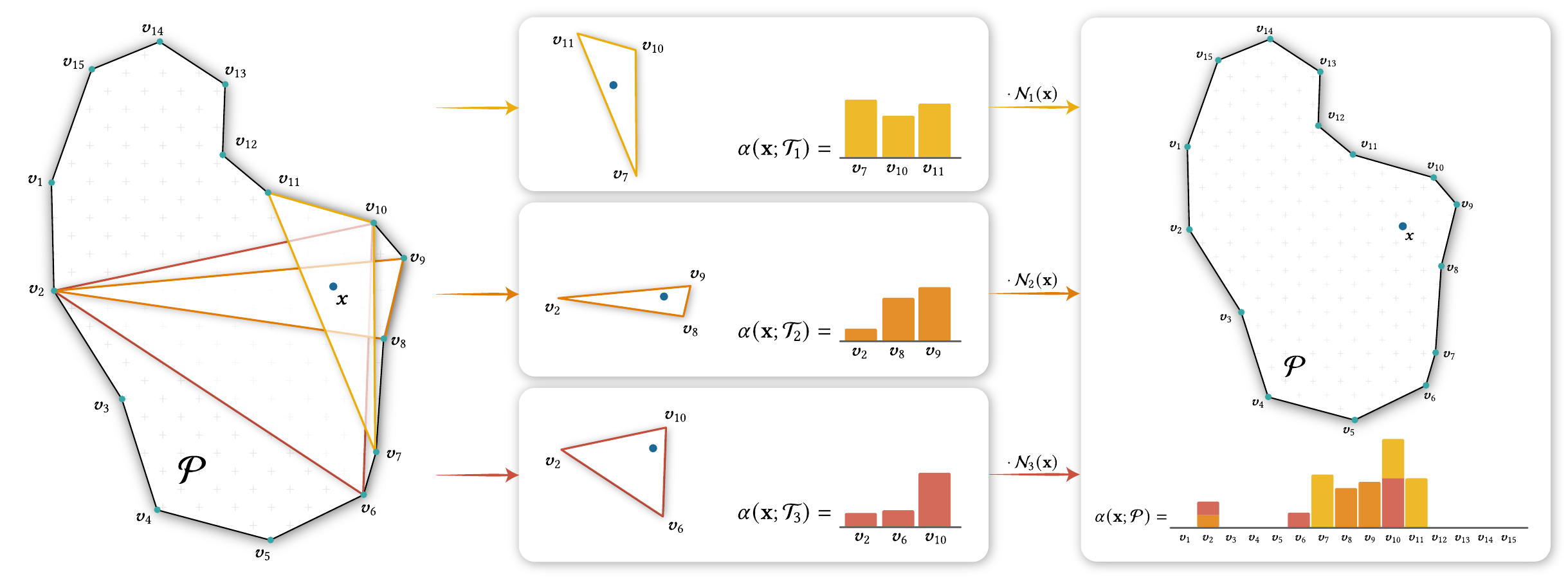}
  \caption{A $2$D illustration of our model. We connect triplets of polygon vertices into non-degenerate triangles. For each triangle that contains $\DomainSample$ in its interior, we compute the triangle barycentric coordinates of $\DomainSample$. We define the polygon barycentric coordinates as the convex combination of triangle barycentric coordinates, with coefficients given by a differentiable parametric function $\mathcal{N}(\DomainSample)$.}
  \Description{An illustration of our model.}\label{fig:explainer}
\end{figure*}

\section{Variational Barycentric Coordinates}

We are now ready to introduce our formulation of generalized barycentric coordinates. We begin by introducing the optimization problem at the heart of our formulation in \S\ref{ss:model} and then offer a computational representation of the function space of valid barycentric coordinates in \S\ref{ss:decomp}.
Additionally, we show that not only are all of the coordinates produced by our method within the feasible set (Proposition~\ref{prop:validity}), but also our method allows us to represent \emph{all possible} valid barycentric coordinates for any cage (Proposition~\ref{prop:univ}).

\subsection{Model}\label{ss:model}

We define variational barycentric coordinates as the minimizer of an energy functional $F$ under the relevant constraints
\begin{subequations}
\setcounter{equation}{-1}%
\renewcommand{\theequation}{\theparentequation}%
\begin{align}
    \min_{\bary_{i \in [1, K]}}
        \sum_{i=1}^{K} \int_\Domain F \left[ \bary_i(\DomainSample), \nabla \bary_i(\DomainSample), \Delta \bary_i(\DomainSample) \right] \Diff V(\DomainSample),%
\end{align}%
\renewcommand{\theequation}{\theparentequation.\arabic{equation}}%
\begin{flalign}%
    && \suchThat & \bary_i(\DomainSample) \geq 0\,,\;\forall \bary_i, \DomainSample, & \eqtag{non-negativity} \label{eq:nonneg} \\[5pt]  
    &&& {\textstyle\sum_i} \bary_i(\DomainSample) = 1\,,\;\forall \DomainSample, & \eqtag{partition of unity} \label{eq:partition} \\[5pt]
    &&& \bary_i(\BoundarySample_j) = \delta_{ij}\,,\;\forall \bary_i, \BoundarySample_j & \eqtag{Lagrange property} \label{eq:lagrange} \\[5pt]
    &&& {\textstyle\sum_i} \bary_i(\DomainSample) \BoundarySample_i = \DomainSample\,,\;\forall \DomainSample, & \eqtag{reproduction} \label{eq:repr}
\end{flalign}\label{eq:model}
\end{subequations}where $\delta_{ij}$ is the Kronecker delta, and $V$ is the volume form of $\Domain$.
Some choices for $F$ have historically included the Dirichlet energy \citet{Joshi2007} and the weighted total variation \citet{Zhang2014}.
Because our method is constrained to the set of valid barycentric coordinates by construction, optimizing for different objectives is simply a matter of using the correct energy functional. 
We will see some other possible choices for $F$ in \S\ref{s:losses}.%

\subsection{Decomposition of Barycentric Coordinates}\label{ss:decomp}

Deep learning architectures provide expressive function approximators, but barycentric coordinates require enforcing the constraints in \eqref{eq:model}---which are not typically satisfied by generic neural network parameterizations.
While constraints in Equations \ref{eq:nonneg}, \ref{eq:partition}, and \ref{eq:lagrange} can be satisfied by normalizing and carefully reweighing the network output, building the reproduction property from Equation~\ref{eq:repr} into a neural network architecture poses a nontrivial challenge. In this section, we will offer a theoretical characterization of barycentric coordinate functions that is amenable to neural network representations.

Recall from \S\ref{s:prelim} that the barycentric coordinates of a simplex in $\R^d$---e.g.\ triangles in $\R^2$, or tetrahedra in $\R^3$---are uniquely determined under mild conditions and can be computed in closed form. 
Due to this fact, our method relies on simplices as its basic building blocks. 
For the sake of simplicity, we will restrict the following discussion to $2$-dimensional polygonal cages. 

Our method decomposes the cage into a large number of non-degenerate overlapping \emph{virtual triangles} that share their vertices with the cage:
\begin{equation}
    \TriSet \coloneq \left\{\Triangle \mid \Boundary(\Triangle) \subseteq \Boundary(\Domain), \text{Vol}(\Triangle) \neq 0 \right\}.
\end{equation}
Let $\Triangle \in \TriSet$ be a virtual triangle with vertices $\left\{ \BoundarySample_l, \BoundarySample_m, \BoundarySample_n \right\} \subseteq \Boundary(\Domain)$. Then, the \emph{cage barycentric coordinates due to $\Triangle$} can be written in terms of the corresponding triangle barycentric coordinates:
\begin{equation}\label{eq:scatter}
    \bary_i(\DomainSample; \Domain, \Triangle) = 
    \begin{cases}
    \bary_{i}(\DomainSample; \Triangle), & \text{if } \BoundarySample_i \in \left\{ \BoundarySample_l, \BoundarySample_m, \BoundarySample_n \right\},\\
    0, & \text{otherwise}.
    \end{cases}
\end{equation}
This formula states that for every $\DomainSample$ inside of $\Triangle$, the triangle  coordinates of $\Triangle$ are also valid cage coordinates as they inherit the necessary constraints.

In practice, multiple triangles typically overlap any given $\DomainSample$.
Therefore, we define the set of \emph{valid triangles at $\DomainSample$} to be all virtual triangles in $\mathbb{T}$ that cover $\DomainSample$:
\begin{equation}
    \TriSet_{\DomainSample} \coloneq \left\{\Triangle \in \TriSet \mid \DomainSample \in \Triangle \right\}.
\end{equation}
To combine the coordinates due to the different virtual triangles, we have to assign each triangle $\Triangle_j\in\TriSet_{\DomainSample}$ a spatially varying weight, $w_{j}(\DomainSample)$. We can then define the full cage barycentric coordinates as a convex combination of the different triangle weights:
\begin{equation}
    \baryv(\DomainSample; \Domain) = \sum_{j = 1}^{\lvert \TriSet_{\DomainSample} \rvert} w_{j}(\DomainSample)\;\baryv(\DomainSample;\Domain, \Triangle_j).
    \end{equation}
As we will show in Proposition~\ref{prop:validity}, so long as $w_j(\DomainSample) \geq 0$ and $\sum_j w_j(\DomainSample)=1$, this operation retains all of the necessary properties such that $\baryv(\DomainSample; \Domain)$ always remain valid barycentric coordinates.

The triangle weights are degrees of freedom, and we are free to choose any set of weights for each $\DomainSample$ such that the weights are valid coefficients of a convex combination.
An equivalent way of phrasing this is that the weights $w_j$ form a \emph{categorical probability distribution} over the elements of $\TriSet_{\DomainSample}$.
Therefore, we can model the weights using any representation that maps points on the interior of the cage to categorical distributions over the set of valid triangles. Luckily, this is a standard problem in machine learning, and we can use a neural network as the triangle weights: ${w_j \coloneq \mathcal{N}_j}$, where ${\mathcal{N}: \R^d \to \Delta(\TriSet_{\DomainSample})}$, and ${\Delta(\TriSet_{\DomainSample})}$ is the probability simplex over $\TriSet_{\DomainSample}$.

This construction leads to parameterization of variational barycentric coordinates:
\begin{flalign}\label{eq:vbc}
  &&&&& \baryv(\DomainSample; \Domain) = \sum_{j = 1}^{\lvert \TriSet_{\DomainSample} \rvert} \mathcal{N}_j(\DomainSample) \;\baryv(\DomainSample; \Domain, \Triangle_j). & \eqtag{VBC}
\end{flalign}
Please refer to Figure~\ref{fig:explainer} for an illustrative explanation and to \S\ref{ss:nn} for more details on the computational model.

As we will see in the proof of Proposition~\ref{prop:validity}, the formula above automatically satisfies \emph{most} of the properties of barycentric coordinates, namely non-negativity, partition of unity, and reproduction. The one missing property is the Lagrange property.  To enforce this property, we have to introduce a slight modification to our formulation.
While there are multiple ways of addressing this challenge, we opted for one that is simple to implement, as it requires no modifications to the architecture, while reducing computational complexity  (see \S\ref{ss:accel}).
In particular, we remove all triangles from $\TriSet$ that contain another cage vertex in their interior.
In other words, we prune triangles containing other vertices of $\Domain$ in their interiors: %
\begin{equation}\label{eq:pruning}
    \widehat{\TriSet} \coloneq \left\{\Triangle \in \TriSet \mid \forall \BoundarySample \in \Boundary(\Domain) : \BoundarySample \in \Triangle \implies \BoundarySample \in \Boundary(\Triangle) \right\}.
\end{equation}
The rest of our formulation remains identical, and we simply operate on $\widehat{\TriSet}$ instead of on $\TriSet$. Now, when we evaluate \eqref{eq:vbc} at $\BoundarySample\in\Boundary(\Domain)$, we are taking a convex combination of barycentric coordinates exclusively from triangles with $\BoundarySample$ as a vertex, all of which fulfill the Lagrange property.

\begin{figure}[t!]
  \centering
  \includegraphics[width=\linewidth]{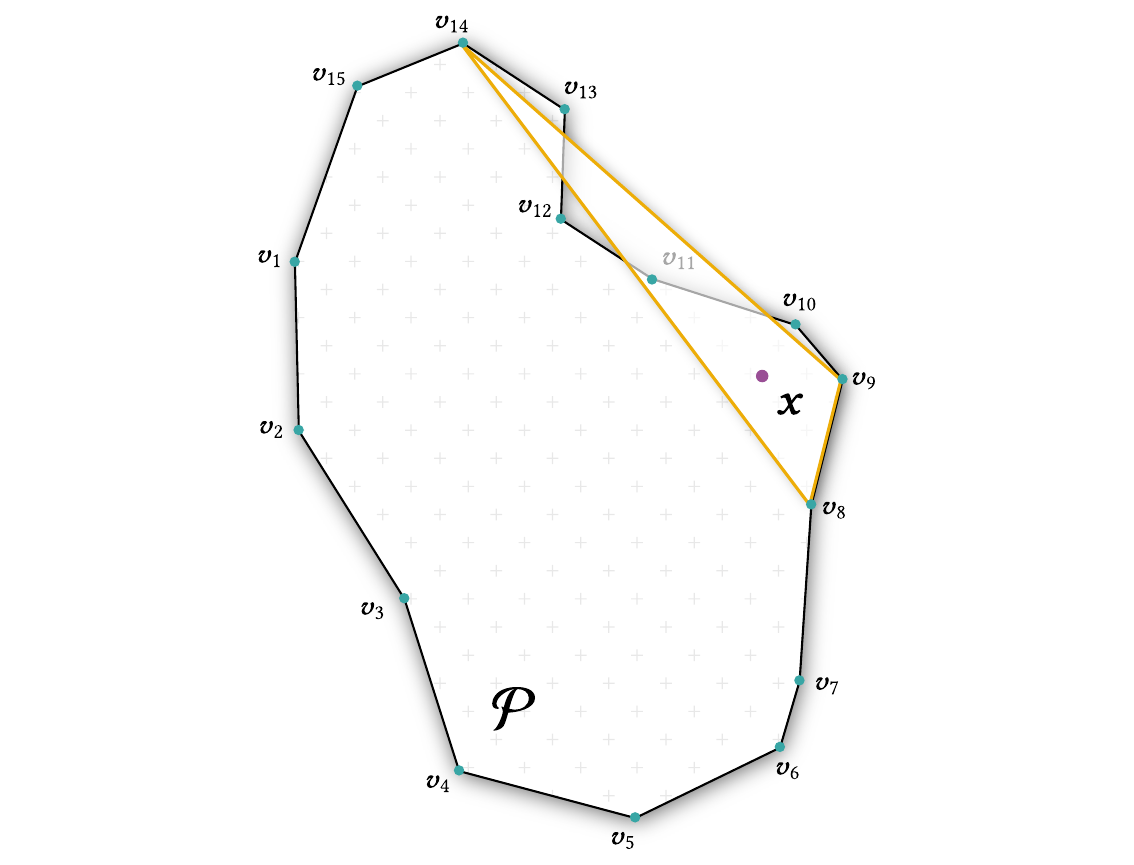}
  \caption{To satisfy the Lagrange property, we prune away triangles which contain polygon vertices in their interior. These triangles introduce long-range dependencies across polygon boundaries, which is considered undesirable \cite{Zhang2014}. The illustrated triangle results in the Lagrange property not being enforced at $\BoundarySample_{11}$, while introducing a long-range dependency between $\DomainSample$ and $\BoundarySample_{14}$. Therefore, pruning these kinds of triangles not only enforces the Lagrange property, but also encourages locality.}
  \Description{Justification for pruning triangles.}\label{fig:prune}
\end{figure}

The proposed modification limits the types of coordinate functions that our model can represent, in contrast with Proposition \ref{prop:univ} below. It would have been possible to solve this problem differently, e.g.\ by reweighting the network outputs to take into account the distance to each cage vertex or by adding a penalty term to the optimization. However, these approaches introduce unnecessary complexity compared to pruning. Moreover, from a practical standpoint, there are few scenarios where having large triangles that cover other vertices is desirable behavior; see Figure~\ref{fig:prune}. Lastly, as we will see in \S\ref{s:comp}, pruning helps in reducing the computational demands of our algorithm.

\begin{figure*}[t!]
  \centering
  \includegraphics[width=\linewidth]{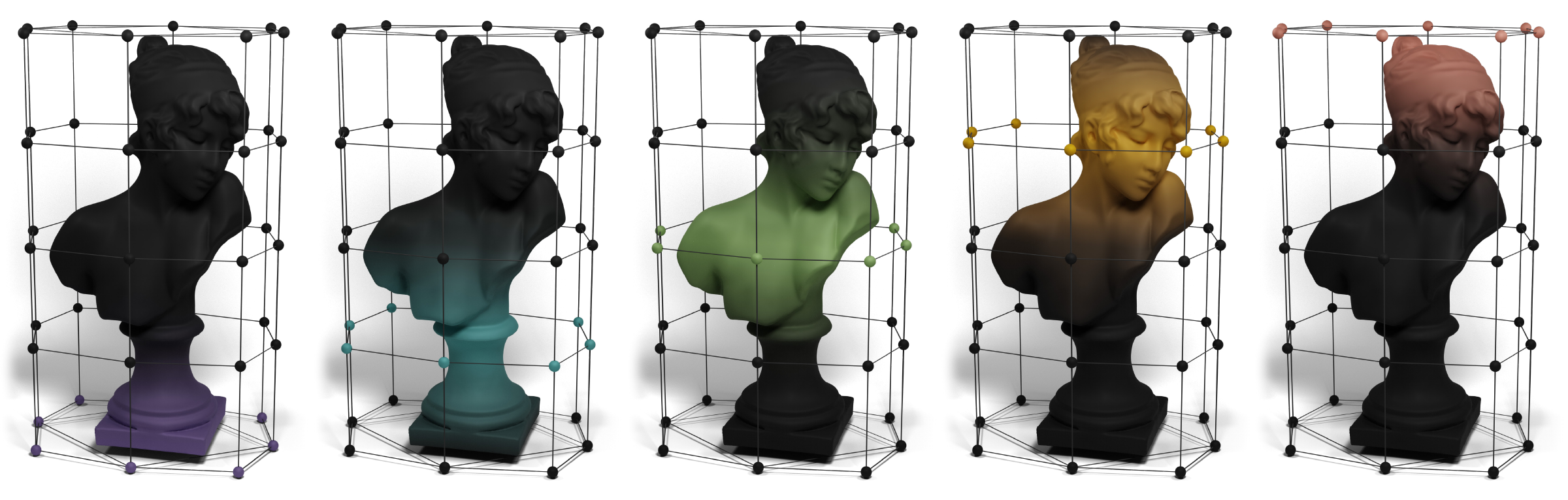}
  \caption{
  Variational barycentric coordinates produce smooth-looking color interpolations from the colored cage vertices onto the \Sappho{} mesh. The coordinates in this Figure were optimized using the weighted TV objective from Section~\ref{ss:wtv}.}
  \Description{Color interpolation using our model.}\label{fig:coords3d}
\end{figure*}

\subsection{Analysis}

Next, we discuss the theoretical properties of our formulation.
In particular, we will show that it produces correct barycentric coordinates by construction, and less obviously, that it can represent \emph{all} barycentric coordinate functions.

\begin{proposition}[Universality]\label{prop:univ}
Given a cage $\Domain$, any barycentric coordinate function %
$\baryv$ %
can be decomposed into a convex combination of simplex barycentric coordinates of all 
possible simplices that can be formed by combining the cage vertices.
\end{proposition}
\begin{proof}
Take a point $\DomainSample\in\Domain$.  By definition, $\baryv(\DomainSample)$ lies in a convex polytope defined by the non-negativity constraint \eqref{eq:nonneg} ($K$ linear  inequality constraints), the partition of unity constraint \eqref{eq:partition} ($1$ linear constraint), and the reproduction constraint \eqref{eq:repr} ($d$ linear constraints).   

The corners of such a polytope cut out by linear inequality constraints are given by intersecting $K$ constraint planes at a time. In this case, there are $d$ reproduction constraints, and there is one partition of unity constraint; this means that each corner intersects at least $K-(d+1)$ non-negativity constraint planes.  This argument shows that $\baryv(\DomainSample)$ contains at most $K-(K-(d+1))=d+1$ nonzero values.  

By the argument above, the corners of the constraint polytope are exactly simplex barycentric coordinates drawn from subsets of vertices of $\Domain$.  Since our constraints are linear, the constraint polytope is contained within the convex hull of its corners, completing the proof.
\end{proof}

A less formal way of phrasing the proposition is that our construction is a \emph{universal barycentric coordinate approximator}. 
Intuitively, we have shown that the constraint polytope of barycentric coordinates is exactly the space in which our network operates, $\Delta(\TriSet_{\DomainSample})$. 
This result shows that our construction captures the fundamental underlying structure of the function space of barycentric coordinates. %
For completeness, we also verify a converse result:
\begin{proposition}[Validity]\label{prop:validity}
As long as the network  $\mathcal{N}(\DomainSample)$'s outputs are non-negative and sum to one, variational barycentric coordinates satisfy the barycentric coordinate constraints by construction.
\end{proposition}
\begin{proof}
Non-negativity and partition of unity are fairly intuitive to verify. Since each set of simplex coordinates $\baryv(\DomainSample; \Domain, \Triangle)$ is non-negative and sums up to one, a convex combination of them will have the same properties.  As discussed near Equation~\ref{eq:pruning}, the Lagrange property follows from an identical argument.

Next, we need to address the reproduction property. We know from Equation~\ref{eq:tribary} that for any individual simplex $\Triangle_j \in \TriSet_{\DomainSample}$, %
\begin{equation}
    \sum_{i = 1}^{d + 1} \bary_i(\DomainSample;\, \Domain, \Triangle_j)\,\BoundarySample_i = \DomainSample.
\end{equation}
Taking a convex combination over the barycentric coordinates of all $\Triangle_j \in \TriSet_{\DomainSample}$ retains the reproduction property,
\begin{equation}
    \sum_{j = 1}^{\lvert \TriSet_{\DomainSample} \rvert} \mathcal{N}_j(\DomainSample) \sum_{i = 1}^{d + 1} \bary_i(\DomainSample; \Domain, \Triangle_j)\,\BoundarySample_i = \sum_{j = 1}^{\lvert \TriSet_{\DomainSample} \rvert} \mathcal{N}_j(\DomainSample)\,\DomainSample = \DomainSample,
\end{equation}
which completes the proof.
\end{proof}

\section{Optimization Objectives}\label{s:losses}

The key benefit of our method compared to previous work is its flexibility with respect to the optimization objective. In this section, we will first discuss how to optimize commonly used first-order smoothness energies: total variation, weighted total variation, and the Dirichlet energy. Additionally, we offer two examples of alternative optimization energies to demonstrate the degree of control over the final look of the animation offered by formulation.

\subsection{Total Variation}

Minimizing total variation (TV) ensures both \emph{piecewise-smoothness} as well as \emph{locality} \cite{Zhang2014, Joshi2007}.
However, our formulation is inherently non-smooth due to the discontinuities along the boundaries of virtual triangles.
As a consequence, minimizing TV requires additional care. Specifically, the familiar definition of total variation \cite{Zhang2014},
\begin{equation} \label{eq:tvstd}
        \text{TV}(\baryv) = \sum_{i=1}^{K} \text{TV}(\bary_i) = \sum_{i=1}^{K} \int_\Domain \left\Vert \nabla \bary_i \right\Vert_2 \Diff \DomainSample,
\end{equation}
is inadequate for our model as it fails to account for the difference in function values across the discontinuities.

Instead, we begin with the more general dual formulation of total variation \cite{ChambolleTV}:
\begin{equation}\label{eq:tvgen}
    \text{TV}(\bary_i) = 
    \sup_{\phi \in C^{\infty}} \left\{
    -\int_\Domain
    \bary_i\, \text{div}\phi\,
    \Diff \DomainSample \Bigm\vert
    \forall \DomainSample \in \Domain :\, \Vert \phi \Vert \leq 1
    \right\}.
\end{equation}
Applying integration by parts to this formulation on a domain with no discontinuities and smooth functions $\bary_i$ recovers Equation~\ref{eq:tvstd}. Following \citet{Paul2020}, we can apply a similar strategy to our problem, taking special care to account for the boundary conditions along discontinuities.

The discontinuities in our formulation partition the domain into $N$ disjoint sets, ${\Domain = \Domain_1 \cup \ldots \cup \Domain_N}$, with a total of $M$ possible discontinuities, ${\partial\Domain_1\,, \ldots \,, \partial\Domain_M}$, between them.
As already alluded to, applying integration by parts to Equation~\ref{eq:tvgen} incurs a boundary term for each discontinuity, $\partial \Domain_j$. Denoting with $\bary_i^{+}$ and $\bary_i^{-}$ the values of $\alpha_i$ on either side of the discontinuity, the final formulation of total variation for our problem is
\begin{equation}\label{eq:tvpart}
    \text{TV}(\bary_i) = %
    \sum_{j = 1}^N
    \int_{\Domain_j}
    \left\Vert \nabla \bary_i \right\Vert_2
    \Diff \DomainSample
    +
    \sum_{j = 1}^M
    \oint_{\partial\Domain_j}
    \lvert \bary_i^{+} - \bary_i^{-} \rvert
    \Diff \DomainSample %
    .
\end{equation}

This version of TV makes it apparent that naively applying automatic differentiation or a finite-difference estimator to the model in Equation~\ref{eq:vbc} would not work, as neither approach correctly captures the boundary integral. At a first glance, it might seem necessary to manually keep track of the discontinuities as well as $\bary_i^{+}$ and $\bary_i^{-}$ during the optimization. However, in \S\ref{ss:smoothing} we will introduce a simple mollification strategy motivated by our total variation formulation in Equation~\ref{eq:tvpart} that will allow us to approximate of this energy with a standard finite-difference estimator.

\subsection{Weighted Total Variation}\label{ss:wtv}

A key component of \emph{local barycentric coordinates} \citep{Zhang2014} is the inclusion of a distance-based weighting of the objective energy,
\begin{equation}\label{eq:weighttv}
    \text{TV}(\baryv) = 
    \sum_{i=1}^{K} \int_\Domain \Psi(d(\DomainSample, \BoundarySample_i))\,\|\nabla \bary_i\|_2 \Diff \DomainSample,
\end{equation}
where $d(\DomainSample, \BoundarySample_i)$ is the geodesic distance between $\DomainSample$ and $\BoundarySample_i$, and $\psi$ is a user-specified function. As a consequence of the acceleration strategy explained in \S\ref{ss:accel}, the geodesic distance to a cage vertex $\BoundarySample_i$ is equivalent to the Euclidean distance in all regions where its coordinates $\bary_i$ are non-zero. This allows us to choose $d(\DomainSample, \BoundarySample_i)$ to be the Euclidean distance between $\DomainSample$ and $\BoundarySample_i$, doing away with geodesic distance solvers and making our method completely grid-free.

In our experiments, we use $\Psi(t) = c + (1-c) t^2$, with $c=10^{-1}$, as a way of blending between standard TV and square-weighted TV. Using only the square-weighted TV would artificially set the smoothness energy near the vertex to close-to-zero. In practice, this means that the optimization procedure would not minimize the discontinuities between virtual triangle edges close to the vertex. 

\subsection{Dirichlet Energy}

The Dirichlet energy was first proposed as an objective for barycentric coordinates by \citet{Joshi2007}. For a smooth function \(u\), it is the integral of its squared gradient over the domain, \(\int \lVert \nabla u \rVert^2 dx\).
Since our functions are not smooth and can be discontinuous over the boundaries \(\partial\Domain_j\), we account for this by introducing additional finite differences over element boundaries to our formulation as a way of mimicking the gradient at these discontinuities. We define the Dirichlet energy for our formulation as
\begin{equation}\label{eq:dirpart}
    \Dir(\bary_i) \coloneqq %
    \sum_{j = 1}^N
    \int_{\Domain_j}
    \left\Vert \nabla \bary_i \right\Vert_2^2
    \Diff \DomainSample
    +
    \sum_{j = 1}^M
    \oint_{\partial\Domain_j}
    \lvert \bary_i^{+} - \bary_i^{-} \rvert^2
    \Diff \DomainSample
    .
\end{equation}
While this is a heuristic approximation of the Dirichlet energy, we find that it yields useful results (Figure~\ref{fig:dirvstv}).
Barycentric coordinates obtained with our formulation of \(\Dir(\bary_i)\) exhibit similar fall-off behavior as Dirichlet coordinates computed by discretizing the entire domain (Figure~\ref{fig:comparison}).

\begin{figure}[h]
  \centering
  \includegraphics[width=\linewidth]{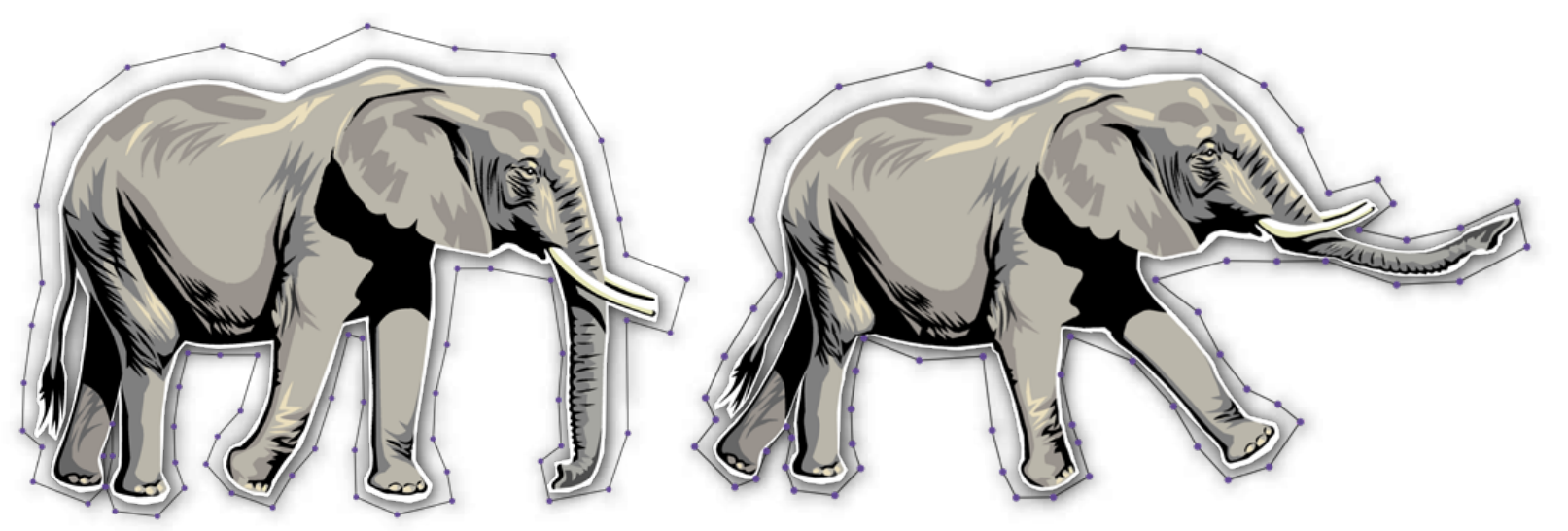}
  \caption{Variational barycentric coordinates produce smooth looking $2$D deformations, as shown on the \Elephant{} mesh. The shown result was obtained by using variational barycentric coordinates optimized using the weighted TV objective from \S\ref{ss:wtv}.}
  \Description{}\label{fig:animation2d}
\end{figure}

\subsection{As-Rigid-As-Possible Energy}\label{ss:arap}

Energies such as total variation or the Dirichlet energy do not account for the final look of the deformations that result from using a set of barycentric coordinates. For example, coordinates produced by minimizing TV often produce elastic-looking deformations, which can be undesirable (see Figures~\ref{fig:teaser} and~\ref{fig:animation2d} for examples). To address this, our approach allows us to fine-tune already optimized weights by using deformation-aware objectives.

In this task, we are given a $3$D model in its rest pose, $\mathcal{S}$, a cage $\Domain$ surrounding that model, and a deformation of the cage, $\Domain'$. The optimization variables are the coordinate functions, $\baryv$ in Equation~\ref{eq:deform}, and the optimization generates weights so that the map $\varphi_{\baryv}$ minimizes the as-rigid-as-possible (ARAP) energy restricted to the surface of the deformed interior mesh, $\mathcal{S}'$ \cite{Sorkine:ARAP, Chao2010, Igarashi:ARAP}.
The continuous ARAP energy is defined as,
\begin{equation}\label{eq:arap}
    \ARAP(\varphi_{\baryv};\;\Domain') = \int_\mathcal{S} \left\Vert d\varphi_{\baryv}(\DomainSample;\;\Domain') - \proj_{\mathrm{SO}(3)} d\varphi_{\baryv}(\DomainSample;\;\Domain') \right\Vert^2_F \Diff \DomainSample,
\end{equation}
where $d\varphi_{\baryv}(\DomainSample;\;\Domain')$ is the push-forward of $\varphi_{\baryv}$ at $\DomainSample$, and $\proj_{\mathrm{SO}(3)}$ the projection onto $\mathrm{SO}(3)$. Following \citet{Sorkine:ARAP}, we have only focused on the surface ARAP energy, i.e.\ the energy of the deformation between $\mathcal{S}$ and $\mathcal{S}^\prime$. We leave volumetric deformation energies \cite{abulnaga2022symmetric} between $\Domain$ and $\Domain'$ for future work.

Since $\mathcal{S}$ is a mesh, we  compute the discrete ARAP energy at each vertex \cite{Sorkine:ARAP} and use it to fine tune a set of variational barycentric coordinates. Specifically, the discrete ARAP energy (assuming uniform Laplacian weights) at an interior vertex $\DomainSample_i$ of $\mathcal{S}$ is defined as
\begin{equation}\label{eq:arapdisc}
    \ARAP_{\DomainSample_i}(\varphi_{\baryv}) \approx \sum_{j \in n(i)} \left\Vert (\varphi_{\baryv}(\DomainSample_i) - \varphi_{\baryv}(\DomainSample_j)) - R_i (\DomainSample_i - \DomainSample_j) \right\Vert^2_F,
\end{equation}
where $n(i)$ is the set of its neighbors, and $R_i$ the best approximating rotation to $d\varphi(\DomainSample_i)$.

During training, we uniformly randomly sample vertices, and, for each sampled vertex, we randomly sample an adjacent edge. A single-sample Monte Carlo estimator of the energy of a given sampled vertex is given by:
\begin{equation}\label{eq:arapmc}
  \widetilde{\ARAP}_{\DomainSample_i}(\varphi_{\baryv}) \approx \frac{1}{|n(i)|} \left\Vert (\varphi_{\baryv}(\DomainSample_i) - \varphi_{\baryv}(\DomainSample_j)) - R_i (\DomainSample_i - \DomainSample_j) \right\Vert^2_F.
\end{equation}
Our final loss is constructed by averaging this single-sample estimator for all sampled vertices.

As a slight modification to the approach by \citet{Sorkine:ARAP}, instead of using SVD to find $R_i$ for each vertex, we store each vertex's rotation as a variable and jointly optimize both $R_i$ and $\varphi_{\baryv}$ using stochastic gradient descent.
In our implementation, we store rotations in the $\mathrm{SO}(3)$ matrix logarithm format. Before running the main optimization loop, we initialize  the rotations by manually finding the best-approximating rotations to the initial $d\varphi_{\baryv}$.

\subsection{Inverse Deformation Energy}\label{ss:inv}

Variational barycentric coordinates can be used to solve inverse deformation problems: given a $3$D model in its rest pose, $\mathcal{S}$, a cage $\Domain$ surrounding that model, and a deformation of the model, $\mathcal{S}'$, we look for a deformation of the cage, $\Domain'$, such that the deformation induced by $\varphi_{\baryv}\left(\mathcal{S};\;\Domain'\right)$ best approximates $\mathcal{S}'$.
Common use-cases are finding the deformation of a human body model to best fit a registered $3$D scan or finding a cage-based deformation that best approximates the result of a physics simulation as a way of amortizing compute. Our model allows us to tackle inverse deformation problems by jointly optimizing our coordinates and the deformed cage vertex positions. We demonstrate the usefulness of VBC for the inverse physics simulation problem in Figure~\ref{fig:inverse} .

Given a set of uniformly randomly drawn samples on the surface of the undeformed mesh, $(\DomainSample_1, \ldots, \DomainSample_N) \in \mathcal{S}$, and the corresponding points on the deformed mesh $(\DomainSample_1^\prime, \ldots, \DomainSample_N^\prime) \in \mathcal{S}'$, we can compute the mean absolute distance to measure the error between the deformation produced by the optimized map and the target mesh.
In addition, we add a regularization term that encourages the norm of the deformed mesh Laplacian to be as close as possible to that of the ground-truth mesh. For each sample $\DomainSample_i$, we compute the norm of the Laplacian at the neighboring triangle vertices and interpolate them onto $\DomainSample_i$. Our inverse deformation energy is given by
\begin{equation}\label{eq:inv}
    \begin{array}{r@{\ }l}
    \text{I}(\mathcal{S}, \mathcal{S}^\prime) =%
    \frac{A(\mathcal{S})}{N} \sum_{i=1}^N %
    &\left\Vert  \varphi_{\baryv}(\DomainSample_i) - \DomainSample_i^\prime \right\Vert_2 %
    \\&+ \lambda \left(\left\Vert \Laplacian \varphi_{\baryv}(\DomainSample_i) \right\Vert_2 - \left\Vert \Laplacian \DomainSample_i^\prime \right\Vert_2\right)^2,
    \end{array}
\end{equation}
where $\Laplacian \DomainSample$ denotes the interpolated vertex Laplacian at $\DomainSample$, $A(\mathcal{S})$ denotes the area of $\mathcal{S}$ and $\lambda$ is a user-specified regularization parameter that we keep fixed at $10^{-4}$.

We optimize Equation~\ref{eq:inv} using a two step procedure. Starting from a weighted TV barycentric coordinate network, we first find the closest-approximating global rotation and scale for the deformed cage using stochastic gradient descent on the energy in Equation~\ref{eq:inv} During, this stage, we keep the deformed cage's local-frame vertex positions and the network weights fixed. Once a suitable global rotation and translation of the deformed cage is found, we jointly optimize the deformed cage's local-frame vertices and fine-tune the barycentric coordinate network. The exact details of the optimization setup for the experiment in Figure~\ref{fig:inverse} are presented in Section~\ref{ss:resdefaware}. 

\section{Computational Model}\label{s:comp}

Having covered the mathematical underpinnings of our model, in this section we detail the practical aspects of its implementation. 

\subsection{Neural Network Model}\label{ss:nn}

We model the convex combinations of triangle barycentric coordinates using a neural field that maps points on the interior, $\DomainSample$ to categorical probability distributions over \emph{valid} virtual triangles in $2$D, i.e.\  tetrahedra in $3$D.
To this end, we construct the last network layer to as many outputs as the \emph{total} number of triangles in $\TriSet$, regardless of whether they contain $\DomainSample$ or not.
The network outputs are mapped to a positive value using the \texttt{softplus} activation function.
We zero out the probabilities for all triangles which do not contain the queried interior point before finally normalizing the distribution.

At first glance, this approach results in a potentially large network output layer, as the total number of virtual simplices scales with $\bigO(K^3)$ in $2$D and $\bigO(K^4)$ in $3$D.
As a remedy, we will introduce a simplex pruning strategy in \S\ref{ss:accel}, which reduces the computational complexity of our method to $\bigO(K)$ for most common scenarios and makes our method tractable for more complex cages.

We use a feed-forward network with $5$ hidden layers of width $256$ and a \texttt{LeakyReLU} \citep{Maas2013RectifierNI} activation function. Before feeding the interior coordinates into the network, we first encode them with the \emph{hash-grid encoding} \cite{mueller2022instant}, with $16$ levels, $4$ features per level, and \texttt{smoothstep} interpolation. We train the network using the Adam optimizer \citet{kingma2014adam}. We include the remaining experimental parameters in Table~\ref{tbl:param}.
We have not made a significant effort to tune the architecture or the hyperparameters of our network, nor have we used any acceleration data structures to accelerate the process of finding which triangles contain a given interior point.

\subsection{Smoothing Discontinuities}\label{ss:smoothing}

Explicitly sampling the discontinuities and computing $\bary_i^{+}$ and $\bary_i^{-}$ in Equation~\ref{eq:tvpart} would add significant additional complexity to our approach.
Instead, we introduce a comparatively simple mollification approach, which allows us to estimate both the interior and the boundary terms in Equations~\ref{eq:tvpart}~and~\ref{eq:dirpart} with an off-the-shelf finite-difference estimator over $\Domain$.

Assume $f$ is zero everywhere on $\R^2$ except within a triangle.
One way of thinking about $f$ is as a smooth function multiplied by the 0-1 indicator function of the triangle.
In essence, our approach replaces the indicator function with an appropriate mollifier, such that the resulting smooth surrogate, $f_{r, \delta}^{*}$, approaches $f$ as $\delta \rightarrow \infty$.
By carefully choosing the smoothing function, we can ensure that $\int \lVert \nabla f^{*}_{r, \delta} \rVert_2$ approaches the TV formulation in Equation~\ref{eq:tvpart}.
Once our model is optimized, we disable the mollification to ensure that the barycentric coordinate constraints are satisfied during inference.

To accomplish this, we define a \emph{smoothing radius} $r$ around each discontinuity, inside of which we rapidly decay the indicator function to zero, and then smooth it using a \emph{scaled logistic} function. Denoting by $d(\DomainSample)$ the signed distance of a point $\DomainSample$ to the boundary of a triangle, we define the a ramp function which increases from linearly from $-1$ to $1$ within the smoothing radius,
\begin{equation}~\label{eq:rampfn}
  R_{r}(\DomainSample) =
  \begin{cases}
      \frac{d(\DomainSample)}{r} & \text{if } d(\DomainSample) \leq | r | , \\
      1 & \text{if } d(\DomainSample) > r, \\
      -1 & \text{if } d(\DomainSample) < -r. \\
  \end{cases}
\end{equation}

\begin{figure}[!t]
  \centering
  \includegraphics[width=\linewidth]{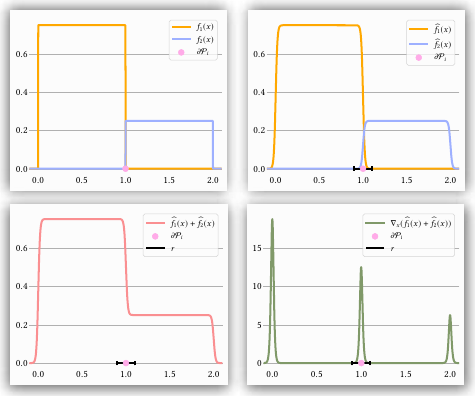}
  \caption{An illustration of our method for computing total variation with a finite difference estimator. The figure in the top left  shows two functions, $f_1$ and $f_2$, which are non-zero over two different line segments (i.e.\ $1$D simplices). The sum of these two functions has a $1$D discontinuity at the vertex that is shared between the two line segments, marked with $\partial \Domain_i$. The top right figure shows the effect of our smoothing term with radius $r$ on each of the two functions. The figure in the bottom left demonstrates that the sum of the individually smoothed functions is equivalent to smoothing the summed function. Lastly, the figure in the bottom right shows the finite differences gradient estimator computed on the smoothed function. We can see that as we increase $r$, the gradient tends to a Dirac delta and the discontinuities are correctly accounted for.}
  \Description{An illustration of how we compute the total variation using finite differences.}\label{fig:smoothing}
\end{figure}

Finally, we smooth the ramp function using a \emph{scaled logistic function}, $\sigma_\delta \left(\ x \right) = \tfrac{1}{1 + \exp\{-\delta x\}}$, where $\delta$ is a user-chosen sharpness parameter,
\begin{equation}
    f^{*}_{r, \delta}(\DomainSample) = 
        \frac{\sigma_\delta \left( R_r(\DomainSample) \right) - 
            \sigma_\delta \left( -1 \right)}{
            \sigma_\delta \left( 1 \right) - 
            \sigma_\delta \left( -1 \right)
        } f(\DomainSample).
\end{equation}
We normalize and recenter $\sigma_{\delta}$ to ensure  continuity at $d(\DomainSample) = \pm r$. %
This function has a $C^1$ discontinuity, which does not appear to affect our optimization procedure in practice; we leave to future work design of a spline-based smoothing function with higher order continuity at ${d(x)=\pm r}$.
Our approach is illustrated in Figure~\ref{fig:smoothing}.

\paragraph*{Choice of smoothing function.} In our scenario, there are multiple indicator functions meeting at the shared edges of virtual triangles. The smoothing function we choose has to have the correct limiting behavior in this situation. The sum of two or more logistic functions centered around the same point is again a logistic function, whose limit converges to a step function. Moreover, the derivative of a logistic function becomes a Dirac delta as $\delta \to \infty$, whose integral equals exactly the difference on either side of the function, as needed to capture the second term in Equation~\ref{eq:tvpart}.

\paragraph*{Implementation details.} Naturally, we only want to perform this type of smoothing on the interior virtual edges, but not at the boundary of $\Domain$. Therefore, it is either necessary to explicitly keep track of which edges are part of the boundary, or to exclude samples closer than $r$ to the boundary from the optimization. We opted for the second approach as it appears not to yield noticeable artifacts as long as $r$ is small enough and the cage faces distant enough from the interior mesh.

There is a natural trade-off when picking the values of $r$ and $\delta$. If $r$ is too small and $\delta$ too large, we might have trouble sampling the discontinuities during optimization. If $\delta$ is too small, we blur the gradient function too much and possibly optimize for the wrong result. The concrete parameters used in our experiments are included in Table~\ref{tbl:param}. %

\paragraph*{Finite-difference estimator.} We use a standard central difference estimator and keep the spacing constant at $h=2.5 \cdot 10^{-2}$. There appears to be a similar trade-off when it comes to the size of $h$ as with the value of $r$: if we make $h$ too small, we amplify the noise due to the network and the hash-grid encoding, but if we make it too large, we increase the bias in the estimated gradient.

\subsection{Accelerating The Model}\label{ss:accel}

So far, the practical utility of our theoretical formulation has been limited by the fact that the number possible virtual triangles grows with $\bigO(K^3)$.
This presents a problem for two reasons---not only does it significantly increase the necessary compute, it also means that the representational power of the neural field quickly becomes insufficient as $K$ becomes larger.
We alleviate both issues by using a simple strategy for pruning triangles, which allows us to limit the number of triangles to a constant multiple of the number of vertices.
Not only does such a strategy make our method tractable, it is also \emph{locality-enforcing}, meaning that it builds a notion locality into our model as a hard constraint.

In \S\ref{ss:decomp}, we presented a simple modification to our formulation that ensures that the necessary constraints are satisfied by eliminating all virtual triangles which contain a cage vertex in their interior. For reasons discussed in Figure~\ref{fig:prune}, we have found this to be a useful heuristic and therefore expand on the idea by completely eliminating all virtual triangles that cross the boundary of the shape. We do this by densely sampling the ambient space around our shape and discarding all virtual triangles that contain one of the samples.

The actual pruning heuristic is designed with locality of coordinates in mind. For each vertex, we first find all virtual triangles that have that vertex as a corner. To enforce locality, we discard triangles whose extent is not limited to the local neighborhood of the vertex. Specifically, we sort the triangles by the length of their longest edge, and keep the $M_p$ shortest ones, where $M_p$ is a user-specified parameter. We will refer to this as the \emph{min-longest-edge} heuristic. Note that this pruning strategy does not replace the one in \S\ref{eq:pruning} and Figure~\ref{fig:prune}. As an example, in cages with three co-linear boundary vertices, we must ensure that no virtual triangle contains the middle of the three boundary vertices on one of its edges.

In certain edge cases, this heuristic can result in parts of the interior not being covered by any virtual triangles. To ensure coverage, we sample the cage interior---either the interior volume, or the surface of an interior mesh---and check if any of the samples have no virtual triangles containing it. If this happens, we proceed to find all virtual triangles that contain the interior point and have the closest cage vertex as a corner, and then use the min-longest-edge heuristic to choose $M_c$ triangles.

We present the steps of our pruning approach in more detail in Algorithm~\ref{alg:pruning}. The parameter values used in our experiments are included in Table~\ref{tbl:param}. We also include relevant statistics in Table~\ref{tbl:qant} to offer a more concrete picture of the computational savings. 

\begin{algorithm}
\caption{Virtual Simplex Pruning}\label{alg:pruning}
\SetKwFunction{Area}{area}
\SetKwFunction{Height}{height}
\SetKwFunction{NotDegenerate}{not\_degenerate}
\SetKwFunction{Any}{any}
\SetKwFunction{Sort}{sort}
\SetKwFunction{MaxEdgeLen}{max\_edge\_length}
\SetKwFunction{RandOutside}{sample\_outside}
\SetKwFunction{RandInside}{sample\_inside}
\SetKwFunction{GetSmallest}{subsample}
\SetKwFunction{ClosestVertex}{closest\_vertex}
\SetKwData{LongestEdge}{"min-longest-edge"}
\SetKwData{IsCovered}{is\_covered}

\SetKwArray{TriSorted}{$\widehat{\TriSet}_{i}^{\text{sort}}$}

\KwIn{Cage $\Domain$, set of all virtual triangles $\TriSet$, no. pruned triangles per cage vertex $M_p$, no. coverage triangles per interior point $M_c$.}
\KwResult{Pruned set of virtual triangles $\widehat{\TriSet}^{\text{prune}}$.}
\;
\Comment{Remove degenerate triangles and triangles not inside $\Domain$:}
$X_{\text{out}} \coloneqq $ \RandOutside{$\Domain$, $N_{\text{out}}$}\;
$\widehat{\TriSet} \coloneqq \{\Triangle \in \TriSet \;\mid\; $\NotDegenerate{$\Triangle$} \And $ (\forall \DomainSample \in X_{\text{out}}: \DomainSample \notin \Triangle) \}$ \;%
\;
\Comment{Prune using the min longest edge heuristic:}
$\widehat{\TriSet}^{\text{pruned}} \coloneqq \emptyset$\;
\For{$\BoundarySample_i$ \KwTo $\Boundary(\Domain)$}{
    \Comment{Find triangles with $\BoundarySample_i$ as a vertex:}
    $\widehat{\TriSet}_i \coloneqq \{\Triangle \in \widehat{\TriSet} \mid \BoundarySample_i \in \Boundary(\Triangle)\}$\;
    
    \Comment{Find $M_p$ triangles with shortest longest edge:}
    $\widehat{\TriSet}^{\text{pruned}} \coloneqq \widehat{\TriSet}^{\text{pruned}} \cup $ \GetSmallest{$\widehat{\TriSet}_i$, $M_p$, \LongestEdge}\;
}
\;
\Comment{Ensure that the interior is covered with triangles:}
$X_{\text{in}} \coloneqq $ \RandInside{$\Domain$, $N_{\text{in}}$}\;
\For{$\DomainSample$ \KwTo $X_{\text{in}}$}{\
    \Comment{Check if $\DomainSample$ is already covered by a triangle:}
    \If{$\{\Triangle \in \widehat{\TriSet}^{\text{pruned}} \mid \DomainSample \in \Triangle\} \neq \emptyset$} {
        \Continue\;
    }
    \Comment{Find triangles that connect to the closest vertex:}
    $\widehat{\TriSet}_i \coloneqq \{\Triangle \in \widehat{\TriSet} \mid $ \ClosestVertex{$\DomainSample$} $\in \Boundary(\Triangle)\}$\;
    $\widehat{\TriSet}^{\text{pruned}} \coloneqq \widehat{\TriSet}^{\text{pruned}} \cup $ \GetSmallest{$\widehat{\TriSet}_i$, $M_c$, \LongestEdge}\;
}
\end{algorithm}

\paragraph*{Discussion.} The coverage step in Algorithm~\ref{alg:pruning} potentially increases the computational complexity beyond a constant factor of the number of cage vertices. However, if the interior mesh is known a priori---which is often the case---we only need to ensure that the vertices of the interior mesh are covered virtual triangles. Doing this reduces the number of virtual triangles down to a constant factor of the sum of cage and interior mesh vertices. We consider this scenario as an edge case as it only occurred in a limited region of one cage mesh from our test data, namely the \Armadillo{} mesh. %

\section{Results}

In this section, we demonstrate the effectiveness and robustness of variational barycentric coordinates. 
First, in \S\ref{ss:val} we present a series of experiments and quantitative results which illustrate the practical behavior of our implementation.
Having validated the method, in \S\ref{ss:qual} we show qualitative results of our method and, in \S\ref{ss:comp}, compare it to previous work.
Lastly, \S\ref{ss:resdefaware} demonstrates the key benefit of our formulation, its ability to optimize for deformation-aware barycentric coordinates.

We built our implementation using PyTorch~\cite{NEURIPS2019_9015}, as well as the Tiny CUDA NN~\cite{tiny-cuda-nn} library. All experiments were performed on a desktop computer with an Intel Xeon E5-2630 v3 CPU, $32$ GB of RAM memory and a Nvidia TITAN Xp GP102 with $12$ GB of VRAM. Note that, despite using the Tiny CUDA NN library, we were not able to take full advantage of it due to the lack of necessary hardware components in our GPU.

\NiceMatrixOptions{cell-space-limits=2pt}
\NiceMatrixOptions{rules/color=[gray]{0.4}}

\begin{table}[H]
\centering
\caption{Experimental model parameters. We use two sets of parameters, one for $2$D experiments, and one for $3$D.}\label{tbl:param}
\begin{NiceTabularX}{\linewidth}{rll}
\CodeBefore
    \rowcolors{2}{gray!10}{}[restart,respect-blocks]
\Body
\toprule
Parameter  & Value ($2$D) & Value ($3$D) \\
\midrule
Smoothing sharpness ($\delta$) & $3000$ & $1000$ \\
Smoothing radius ($r$) & $5 \cdot 10^{-3}$ & $8 \cdot 10^{-3}$ \\
Max. simplices per cage vertex ($M_p$) & $28$ & $80$ \\
Max. simplices per interior point ($M_c$) & $5$ & $5$\\
\midrule
Training steps & $2000$ & $3000$ \\
Learning rate & $10^{-3}$ &  $5\cdot10^{-4}$ \\
Batch size & $3000$ & $2000$ \\
\bottomrule
\end{NiceTabularX}
\end{table}

\begin{table*}[t!]
\begin{NiceTabularX}{\textwidth}{rX[l]lllllllll}
\CodeBefore
    \rowcolors{2}{gray!10}{}[restart,respect-blocks]
\Body
\toprule
 && \Star{} & \Gecko{} & \Woody{} & \Elephant{} & \BlueMonster{} & \Horse{} & \Sappho{} & \Hand{} & \Armadillo{} \\
\midrule
Cage vertices && $12$ & $34$ & $26$ & $64$ & $100$ & $51$ & $40$ & $92$ & $110$ \\
Possible simplices && $220$ & $5984$ & $2600$ & $41664$ & $16170$  & $249900$ & $91390$ & $2794155$ & $5773185$ \\
Interior simplices && $68$ & $388$ & $738$ & $1923$ & $5255$ & $3247$ &  $83958$ & $35462$ & $56438$ \\
Used simplices && $68$ & $388$ & $403$ & $803$ & $1469$ & $1072$ & $1308$ & $3432$ & $4153$ \\
\cmidrule{1-11}
Training time (min) && $1.03$ & $1.36$ & $1.37$ & $1.99$ & $3.09$ & $11.88$ & $15.45$ & $35.32$ & $43.05$ \\
Inference time (ms) && $3.64$ & $4.49$ & $4.66$ & $7.17$ & $11.41$ & $42.70$ & $56.78$ & $136.26$ & $169.66$ \\
\bottomrule
\end{NiceTabularX}
\caption{Experimental statistics. As a way of quantitatively characterizing our pruning heuristic, we include in the table for a subset of cage meshes the number of vertices, the total number of possible virtual simplices, the number of non-degenerate simplices fully contained in the interior of the cage, as well as the final number of simplices used in the model. We also include the training times, as well as the inference times for a batch size of $2000$.}\label{tbl:qant}
\end{table*}

\begin{figure}[h]
  \centering
  \includegraphics[width=\linewidth]{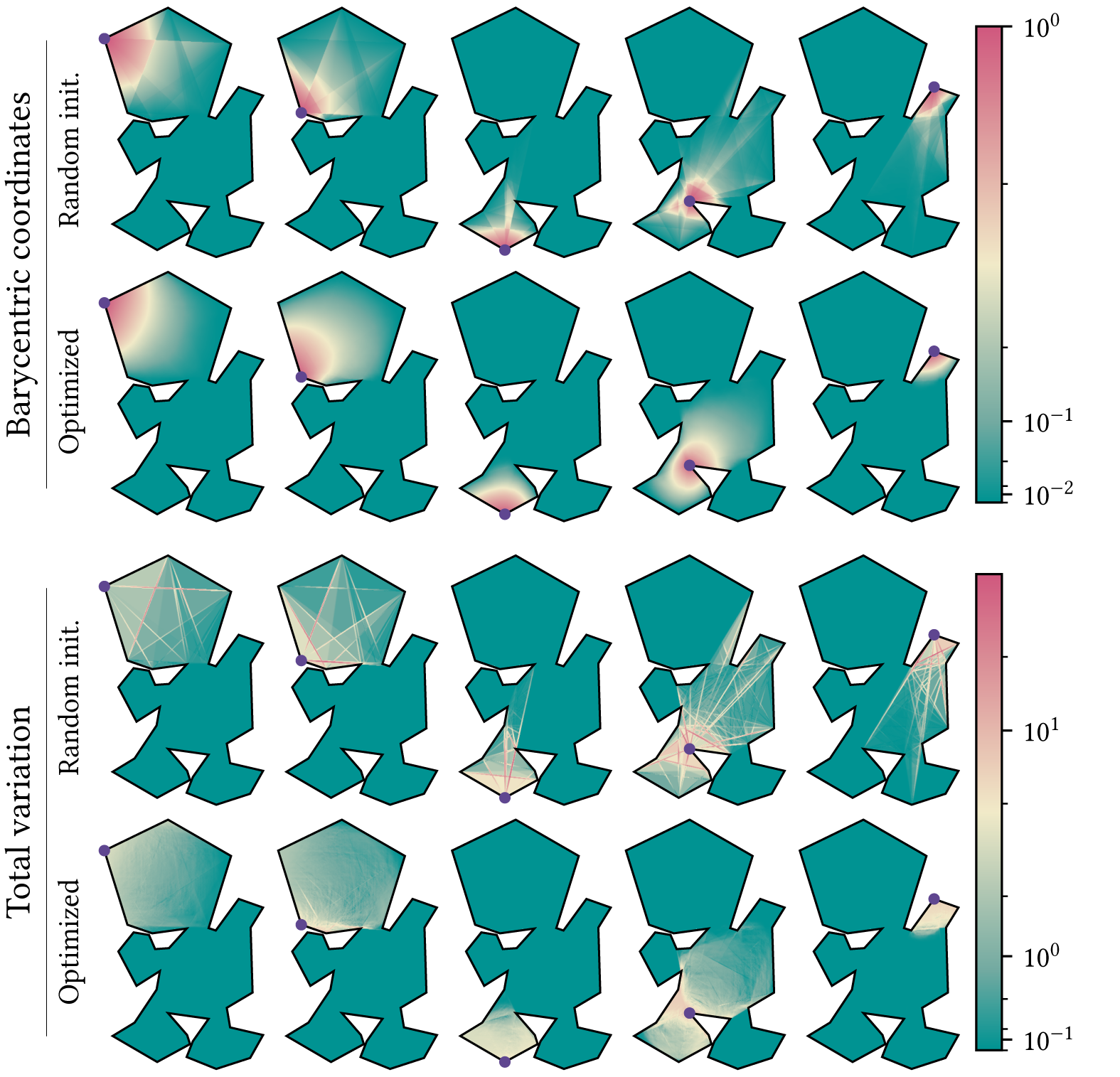}
  \caption{Visualizing the $2$D variational barycentric coordinates as well as the associated TV energy before and after optimization on the \Gecko{} mesh.}
  \Description{Visualizing the 2D variational barycentric coordinates and TV.}\label{fig:coords2d}
\end{figure}

\subsection{Validation}\label{ss:val} %

In this section, we discuss the different practical outcomes of our implementation.  We scale each cage mesh to the unit square, (resp. unit cube) while maintaining the aspect ratio of its bounding box. This allows us to have only two sets of parameters across all of our test meshes: one for $2$D and one for $3$D.

To determine suitable parameters for the discontinuity smoothing in \S\ref{ss:smoothing}, as well as the number of virtual simplices $M_p$, we performed two ablation studies. An excerpt of the results is shown in Figure~\ref{fig:ablation}, with the entire ablation study available in the supplemental material. In summary, we want to make $\delta$ as large as possible to best approximate Equation~\ref{eq:tvpart}, while still keeping it small enough as to not cause numerical issues. As for the number of virtual simplices, we require a large enough value of $M_p$ for sufficient representation power. Interestingly, the quality seems to plateau after a certain number of virtual triangles, practically justifying the min-longest-edge-heuristic. Table~\ref{tbl:param} contains the concrete algorithm parameters we used for all experiments.

Each $3$D training sample has a lower probability of being in the smoothing radius of each discontinuity, making the training signal in those areas sparser and more noisy.
Therefore, we found it useful to use a wider smoothing function in $3$D compared to $2$D.
In general, we also require more virtual simplices per cage vertex in $3$D compared to $2$D to produce smooth coordinate function.
This is to be expected, as there are far more possible virtual tetrahedra than there are triangles.
The parameter which governs the number of simplices per interior point in the scenario where our initial pruning heuristic fails to cover the entire domain, $M_c$, remained unused in all of our $2$D experiments.
We nonetheless suggest leaving it as a non-zero value to ensure against corner cases.

\begin{figure}[h]
  \centering
  \includegraphics[width=\linewidth]{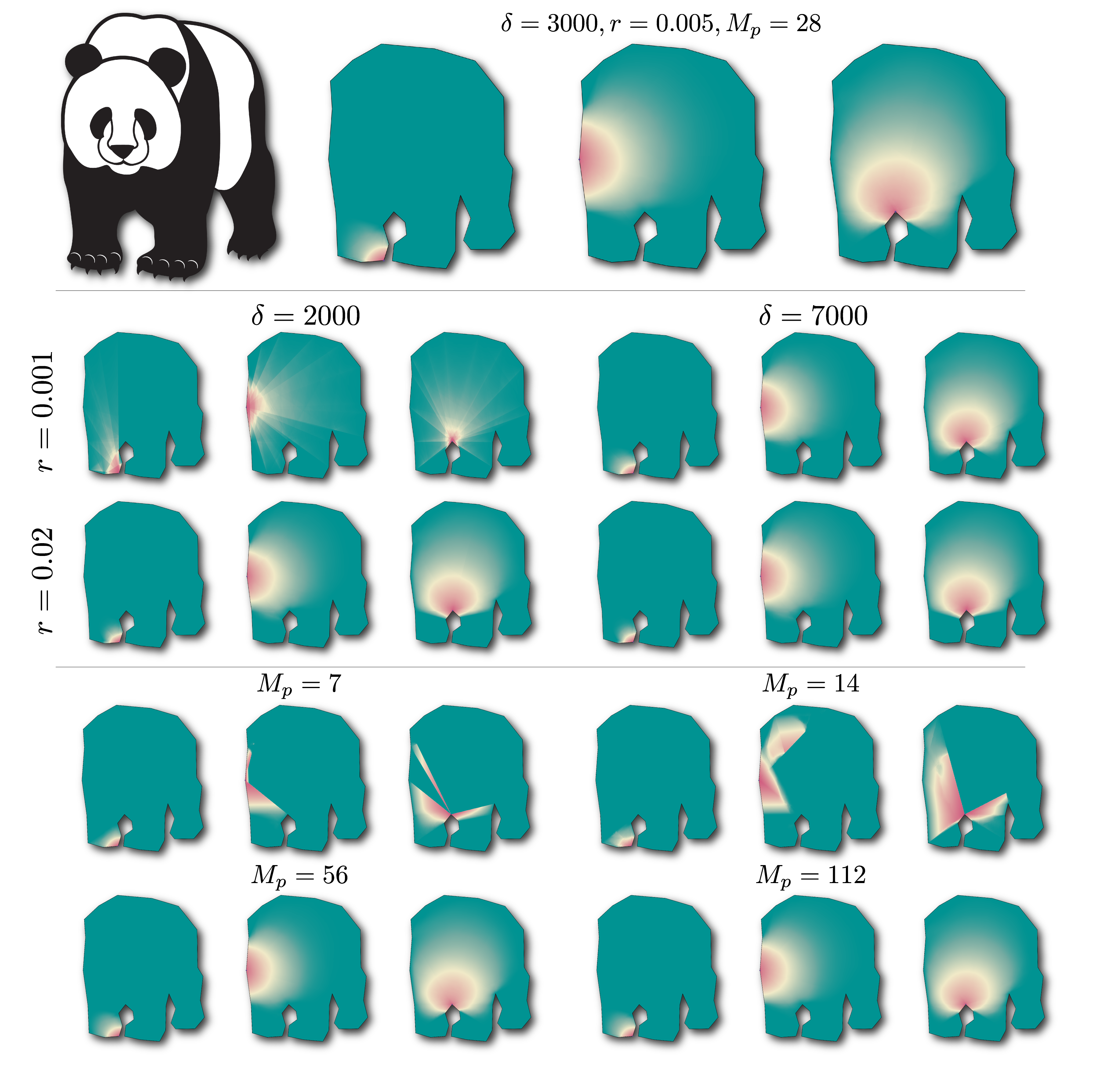}
  \caption{A subset of the results on the \Panda{} mesh from our ablation study. We find larger values of the sharpness parameter $\delta$ to perform better as they help better approximate Equation~\ref{eq:tvpart}. The visual smoothness of the results deteriorates again after a certain point, presumably due to numerical issues caused by large gradient values associated with larger values of $\delta$. Making the smoothing radius $r$ too small compared to the sharpness parameter $\delta$ results in visual artifacts. We hypothesize that this is due to discontinuities caused by the ramp function cutoff in Equation~\ref{eq:rampfn} which otherwise become numerically negligable when $\delta$ is large enough compared to $r$. For the ablation study of the min-longest-edge heuristic, we set $M_c=0$, disabling the part of the algorithm that ensures interior coverage, and varied the maximum number of virtual triangles per cage vertex, $M_p$. As expected, small values of $M_p$ are not enough to faithfully represent the coordinate functions in the entire shape, and after a certain point, adding additional virtual triangles does not affect the results visually.}
  \Description{...}\label{fig:ablation}
\end{figure}

Table~\ref{tbl:qant} presents the quantitative behavior of our algorithm, including statistics related to our pruning heuristic from Section~\ref{ss:accel}.
While the number of all possible simplices increases with the number of cage vertices and the dimension of ambient space, our heuristics keep the final number of simplices in our model to a more reasonable number.

While the training and inference time increase somewhat with the number of cage vertices, the main factor influencing timing is dimension ($2$D vs.\ $3$D). 
We attribute this to several factors: the inference time is slower due to a larger final network layer, a naive $3$D central difference estimator requires $2$ additional network evaluations compared to a $2$D one, and we require a smaller batch size with more training steps and a smaller learning rate to accomodate the increased GPU memory requirements due to the two additional network evaluations each training step.

Figure~\ref{fig:coords2d} visually compares the variational barycentric coordinates and associated TV energy of a randomly initialized network to those of a TV-optimized network.
The total variation of the randomly initialized network is focused primarily on the discontinuities between virtual triangles.
These are largely optimized-away, demonstrating the effectiveness of the smoothing approach from \S\ref{ss:smoothing}.

In this experiment, the \emph{full} total variation, $TV(\baryv) = \sum_i TV(\bary_i)$, of the optimized network is non-zero on the entire domain.
This demonstrates a tension between barycentric coordinate constraints and smoothness energies: 
Because all barycentric coordinate functions are convex combinations of simplex barycentric coordinates, their gradients are also complex combinations of the corresponding simplex gradients towards the cage vertex.

\subsection{Qualitative Results}\label{ss:qual}

\begin{figure}[t]
  \centering
  \includegraphics[width=\linewidth]{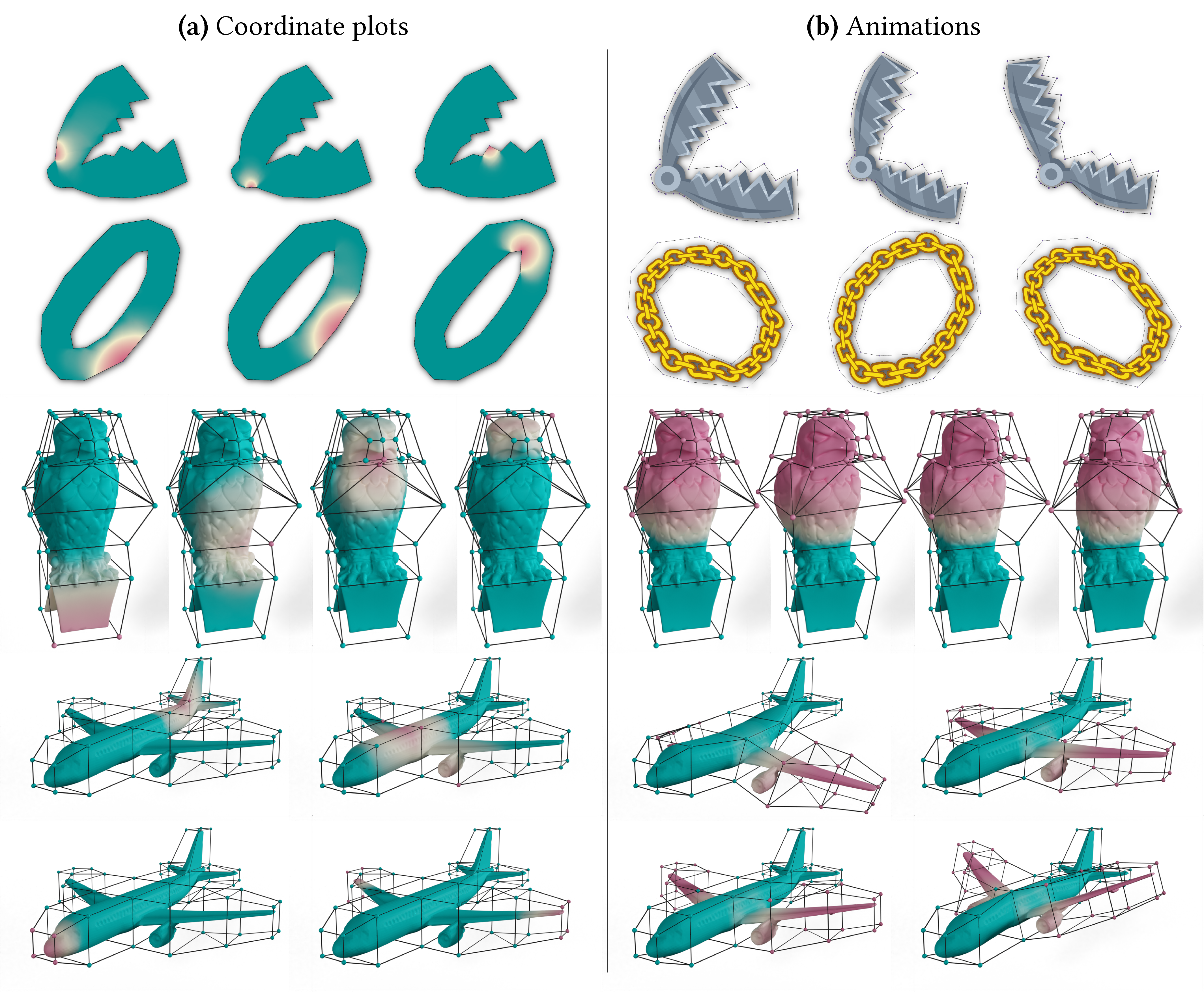}
  \caption{Figure (a) shows the visualization of weighted TV coordinates on $2$D and $3$D meshes. Figure (b) shows excerpts from smooth animations generated by inbetweening cage deformations.}
  \Description{...}\label{fig:coordsandanim}
\end{figure}

In this section, we provide several visualizations of variational barycentric coordinates in both 2D and 3D and demonstrate their usefulness for cage-based deformation.

\begin{figure}[h]
  \centering
  \includegraphics[width=\linewidth]{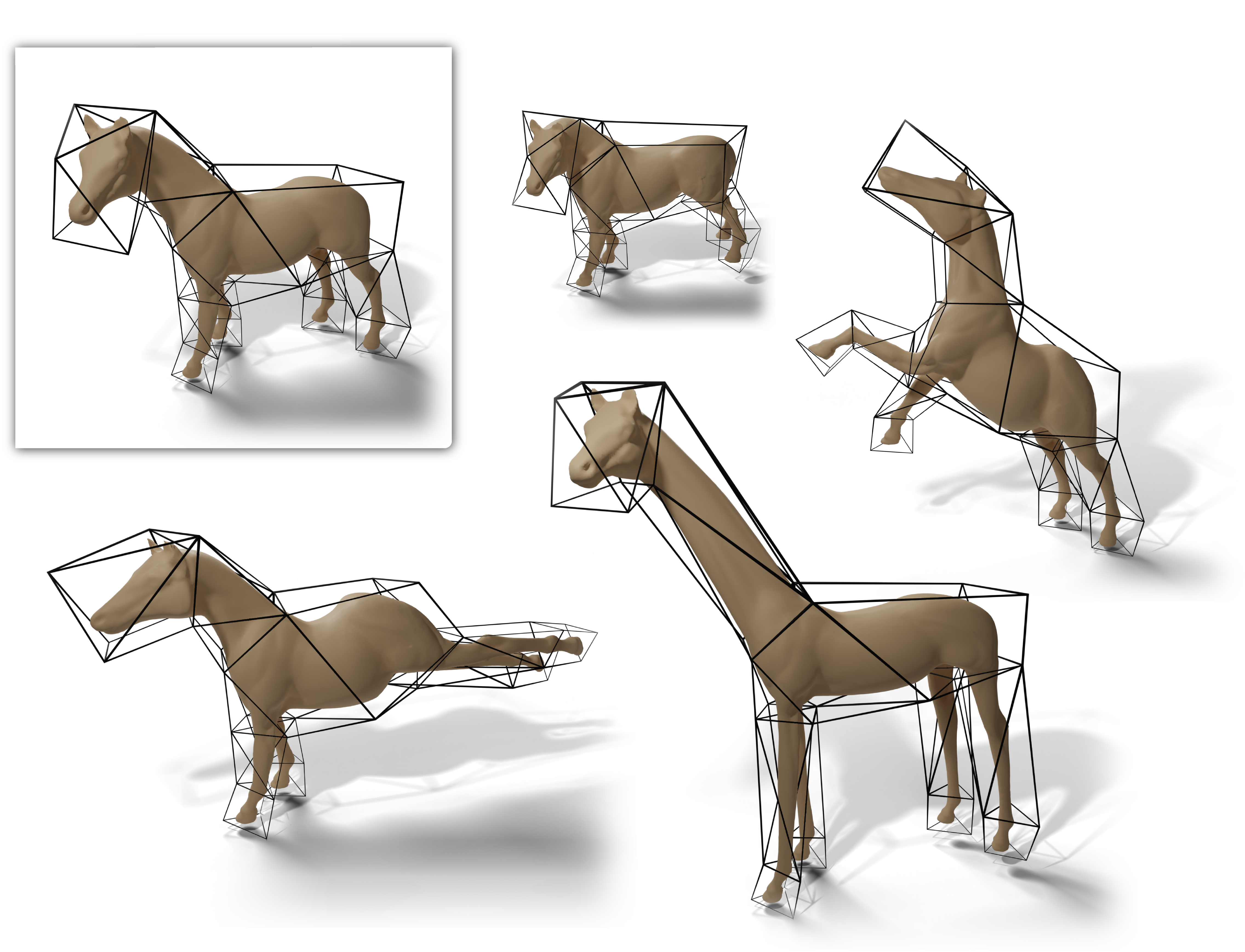}
  \caption{We demonstrate the effect of using TV-optimized variaitonal barycentric coordinates to perform various deformation of the \Horse{} mesh. The undeformed mesh is shown on the top left.}
  \Description{}\label{fig:animation3d}
\end{figure}

Figure~\ref{fig:coords3d} demonstrates how our coordinates can be used to smoothly interpolate values such as color from the vertices of a $3$D cage into its interior.
Similar to the $2$D example in Figure~\ref{fig:coords2d}, the TV variational barycentric coordinates produce smooth-looking function interpolations.

\begin{figure}[h]
  \centering
  \includegraphics[width=\linewidth]{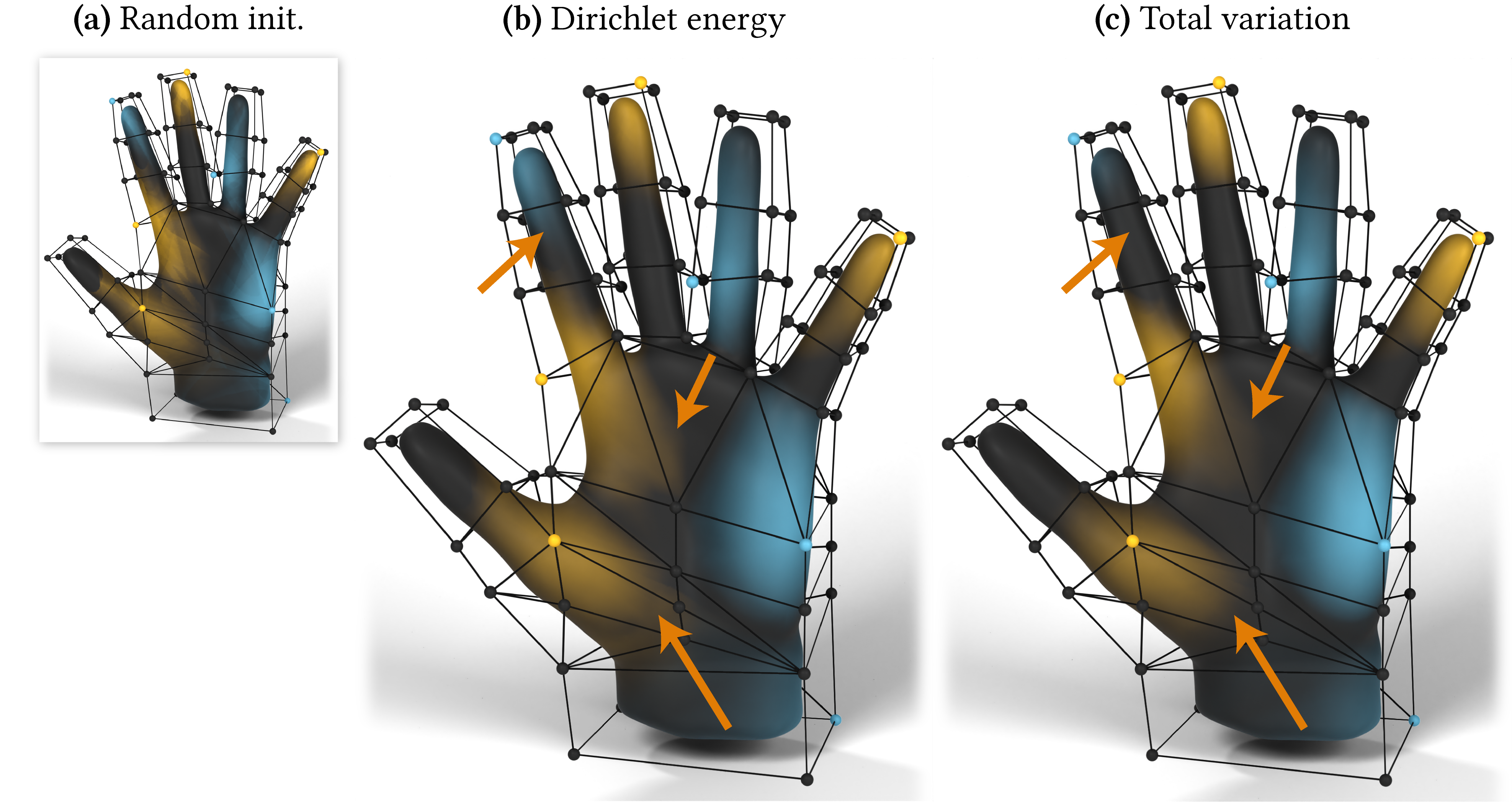}
  \caption{Comparing different sets of barycentric coordinate functions in 3D. Left to right, we show coordinates from a randomly initialized network prior to training, our Dirichlet coordinates, as well as our weighted TV coordinates. The Dirichlet-optimized coordinate functions tend to be blurrier and have a wider region of influence compared to the TV-optimized ones.}
  \Description{...}\label{fig:dirvstv}
\end{figure}

Figure~\ref{fig:dirvstv} depicts an experiment where we compare the barycentric coordinate functions obtained by minimizing the Dirichlet energy as opposed to total variation.
The total variation coordinates tend to be significantly more localized compared to the Dirichlet ones, leading to narrower regions of influence.
The flexibility of our formulation with respect to the optimization objective allows the user to choose the set of coordinate functions which best suits their particular use-case.

Lastly, Figures~\ref{fig:animation2d},~\ref{fig:coordsandanim}~and~\ref{fig:animation3d} demonstrate the usefulness of our coordinates for the common use-case of $2$D and $3$D cage-based deformation. As evidenced by our experiments, the deformations appear smooth, and the deformations due to individual cage vertices do not exhibit unnecessarily global influence, in part due to the pruning heuristic from Section~\ref{ss:accel}. We refer the reader to the supplemental material for the animation videos.

\begin{figure*}[t!]
  \centering
  \includegraphics[width=\linewidth]{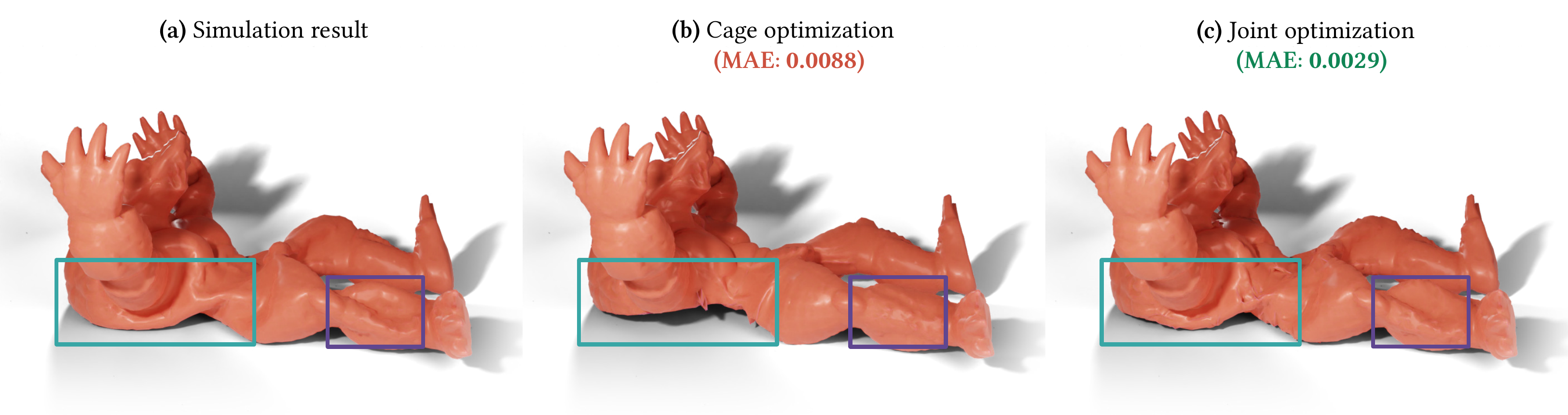}
  \caption{Our method can be used to approximate non-linear deformations such as the soft-body simulation result shown in Figure~(a). Simply using the standard TV coordinate and optimizing the cage leads to an unnatural result, visible artifacts and a large per-vertex approximation error, as visible in Figure~(b). However, if we also allow the weights of our network to change during the optimization, we can achieve a much more natural looking result, decreasing the error by two thirds---see Figure~(c).}
  \Description{...}\label{fig:inverse}
\end{figure*}

\subsection{Comparison}\label{ss:comp}

\begin{figure}[h]
  \centering
  \includegraphics[width=\linewidth]{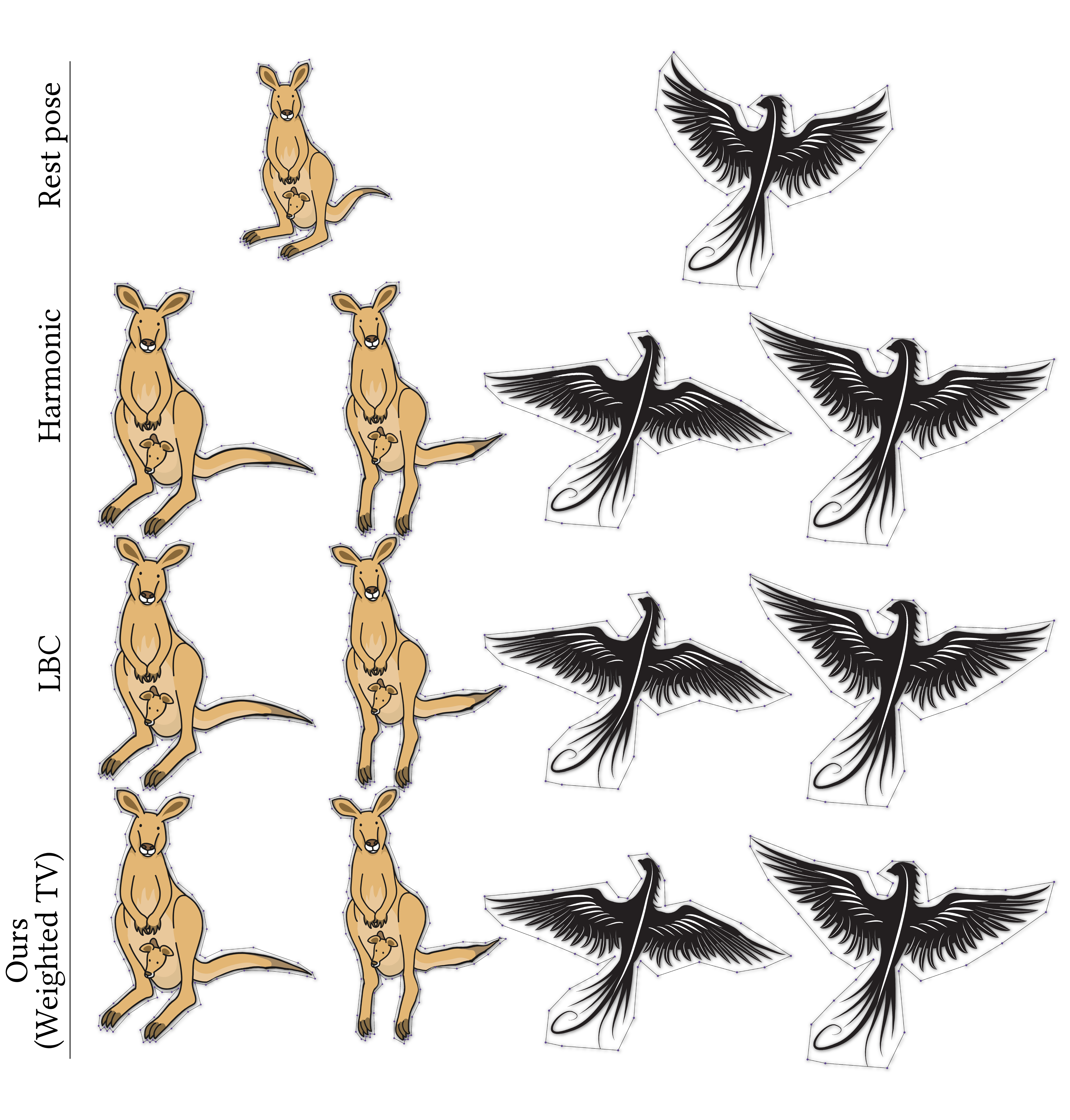}
  \caption{Select frames taken from smooth animations produced by our method compared to those produced by previous work (full videos available in the supplemental material).}
  \Description{...}\label{fig:defcomp}
\end{figure}

\begin{figure*}
\centering
\includegraphics[width=\linewidth]{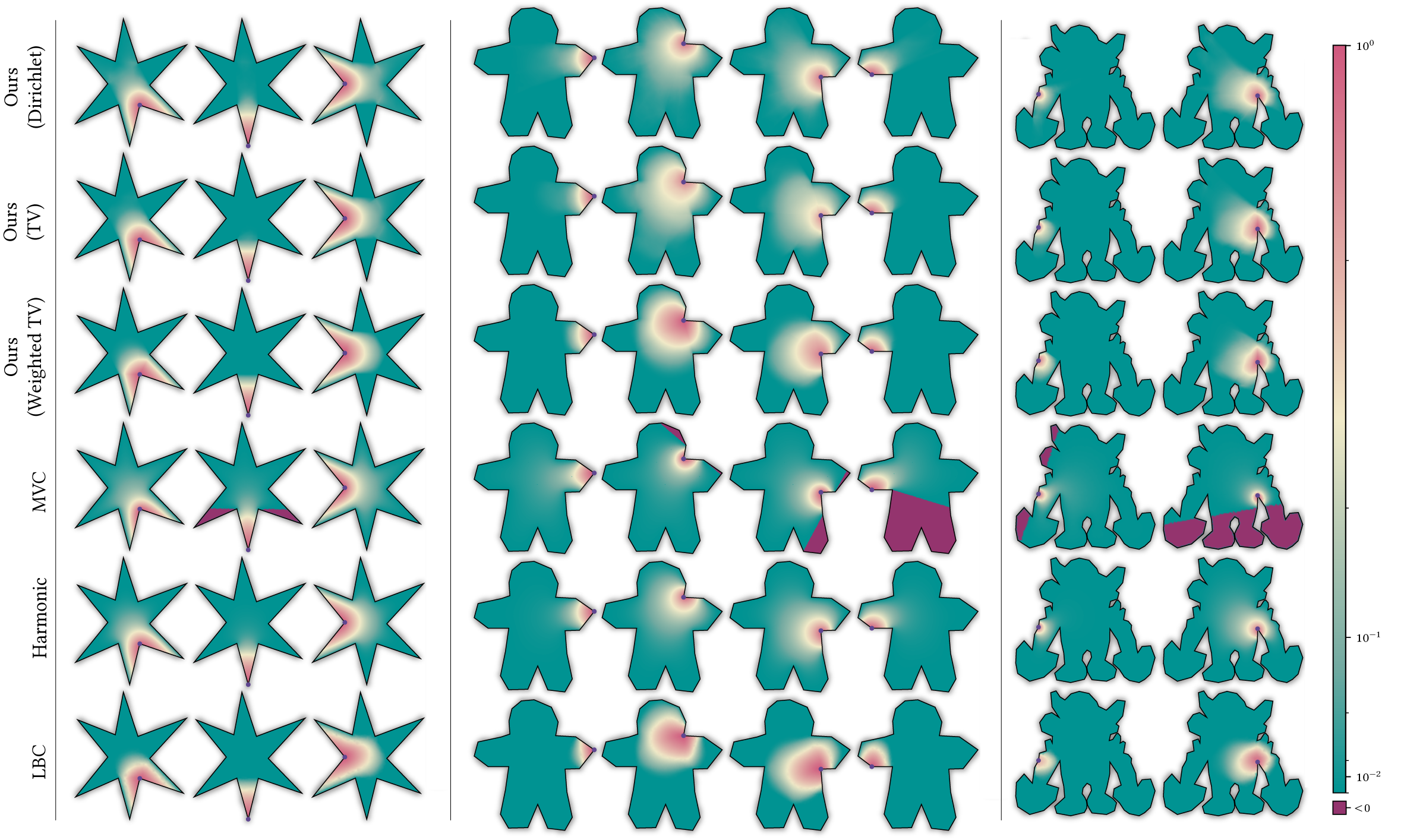}
\caption{Our Dirichlet, TV, and weighted TV coordinate functions, compared to the basis functions produced by Mean Value Coordinates \cite{Floater2003, Ju2005}, Harmonic Coordinates \cite{Joshi2007}, and Local Barycentric Coordinates \cite{Zhang2014} and  on three different shape.
\Description{Descr}}\label{fig:comparison}
\end{figure*}

Figures~\ref{fig:comparison}~and~\ref{fig:defcomp} compare our Dirichlet, TV, and weighted TV coordinates with 
other available methods.
All produce broadly similar weights that fulfill that barycentric conditions, but differences exist.
Our weighted TV results are qualitatively similar to Local Barycentric Coordinates (LBC), which also uses a weighted TV energy.
Weighted TV coordinates have more local support (no small values far away from the cage vertex), and a harsher fall-off of basis functions (the area of large values near the vage vertex is larger).
This effect is slightly more pronounced for our weighted TV results compared to LBC's.
Moreover, unlike LBC, our TV and weighted TV methods do not require discretization of the domain.
MVC generates particularly nonlocal results: while the coordinates' red values in our weighted TV function can be seen to almost partition the shape, MVC coordinates are red only extremely close to the cage vertex, and all blend into each other in the interior of the shape.
The Dirichlet and MVC coordinate functions have rounder isolines than the TV-based methods.
This is a consequence of the order of the exponent in the energy's norm.
Our Dirichlet coordinates exhibit maximum overall smoothness and smaller areas with large function values, while having larger supports, without the need to discretize the shape's interior.
Our non-weighted TV coordinates' behavior is in between the Dirichlet and weighted TV coordinate functions.
The resulting coordinates are not as local as the weighted TV coordinates, but they are smoother, and they are more local than the Dirichlet coordinates, but not as smooth.

\subsection{Deformation-Aware Coordinates}\label{ss:resdefaware}

A distinguishing feature of our formulation is its ability to optimize different objective functions within the space of barycentric coordinate functions. Already, in Section~\ref{ss:qual}, we demonstrated several results with two such functions: the total variation from Equations~\ref{eq:tvpart}~and~\ref{eq:weighttv}, and the Dirichlet energy from Equation~\ref{eq:dirpart}. Beyond these two generic energies, in this section we test the \emph{as-rigid-as-possible} energy introduced in Section~\ref{ss:arap}, as well as the inverse deformation energy from Section~\ref{ss:inv}.

\begin{figure}[h]
  \centering
  \includegraphics[width=\linewidth]{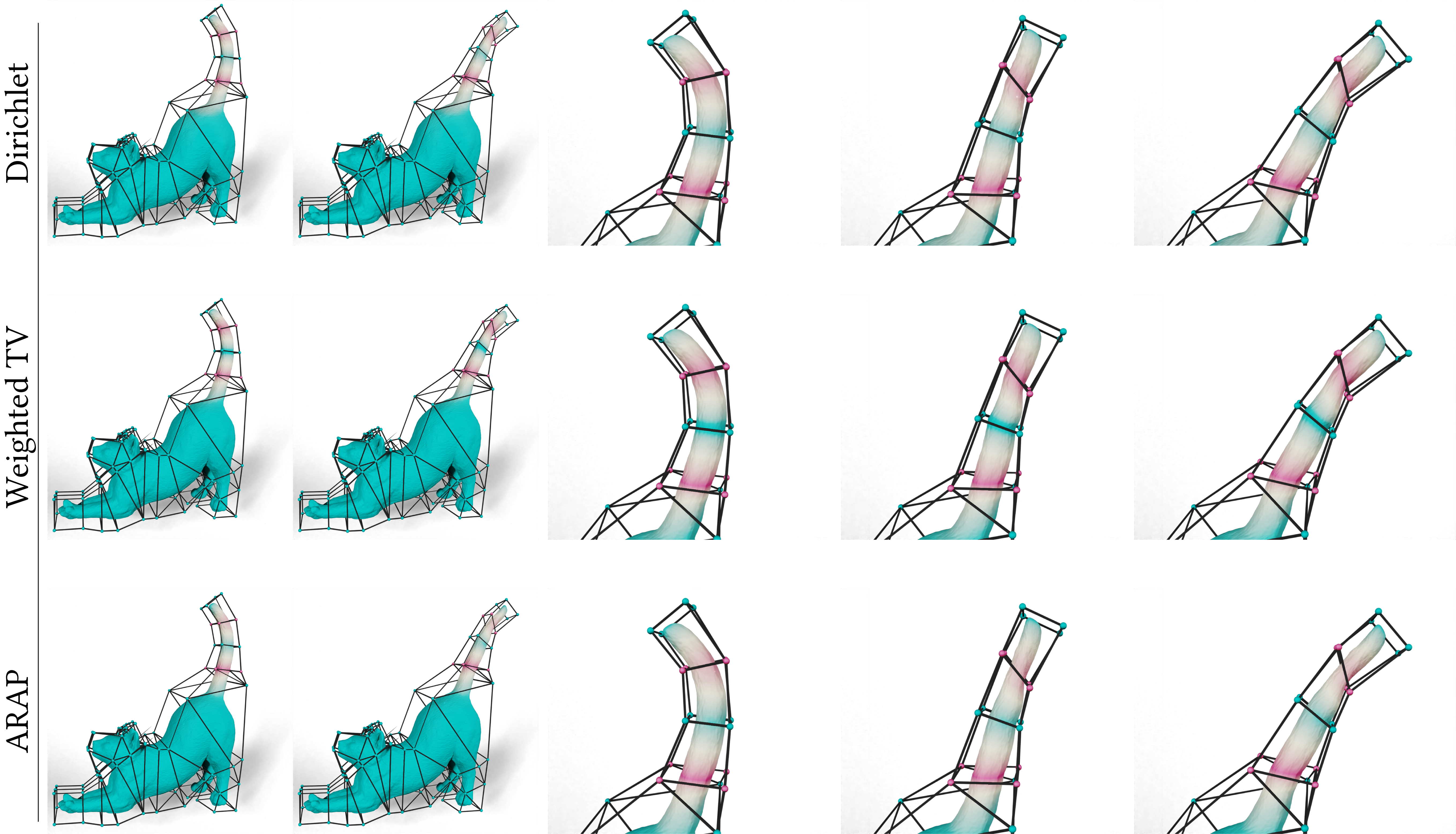}
  \caption{Visualizing the $3$D variational barycentric coordinates during an animation generated by inbetweening deformations. The Dirichlet coordinates have a more non-local influence compared to the Weighted TV objective, resulting in less abrupt bends near the tip of the tail. Nonetheless, we are able to make the Weighted TV deformation more rigid looking by minimizing the ARAP objective. Note that, even though the distortion was minimized only between the undeformed shape and the shape at the $36$\textsuperscript{th} frame, the ARAP optimization improves the look of the entire animation.}
  \Description{}\label{fig:catarap}
\end{figure}

\paragraph*{As-rigid-as-possible coordinates} Figures~\ref{fig:teaser}~and~\ref{fig:catarap} show examples of using the ARAP energy from Section~\ref{ss:arap} to optimize for coordinates that produce less elastic-looking deformations.
In Figure~\ref{fig:teaser}, the initial result was obtained by using weighted TV coordinates to deform the \Armadillo{} mesh. Fine-tuning the coordinates with the surface ARAP energy resulted in fewer artifacts and more rigid-looking deformations, as evidenced by the highlighted regions of the figure.

Figure~\ref{fig:catarap} demonstrates how the ARAP energy can be useful in producing rigid and smooth looking deformations (refer to the supplemental material for videos). To generate the animation, we manually deformed the cage at a set of key-frames and inbetweened the cage vertices for the rest of the frames. We used the key-frame with the most extreme deformation of the cage as $\Domain'$ when optimizing the energy in Equation~\ref{eq:arapdisc}. Nonetheless, we see that the entire animation is affected by the ARAP coordinates, making it look overall more rigid.

In our implementation, we found it necessary to use a significantly larger learning rate for the parameters of the local rotations compared to the network weights ($0.1$ and $10^{-3}$, respectively). We trained the model for a total of $1200$ steps, decaying the learning rate by a factor of $0.8$ every $150$ steps. 

\paragraph*{Inverse deformation} Our model is fully differentiable, allowing us to solve inverse deformation problems. In  Figure~\ref{fig:inverse} we created a high-resolution soft-body physics simulation in Blender to use as the target for optimization. As shown by our results, optimizing the positions of the cage vertices alone is insufficient to faithfully reproduce the simulation result. By comparison, jointly optimizing the weights of the network produces fewer artifacts and achieves a significantly lower reproduction error. Once optimized, these coordinates could be used to transfer the deformations due to a physics simulation from low-resolution meshes onto high-resolution ones similar to \citet{Sacht:2015:NC} or onto the same mesh after an editing operation that changed the mesh topology. 

\begin{figure}[h]
  \centering
  \includegraphics[width=\linewidth]{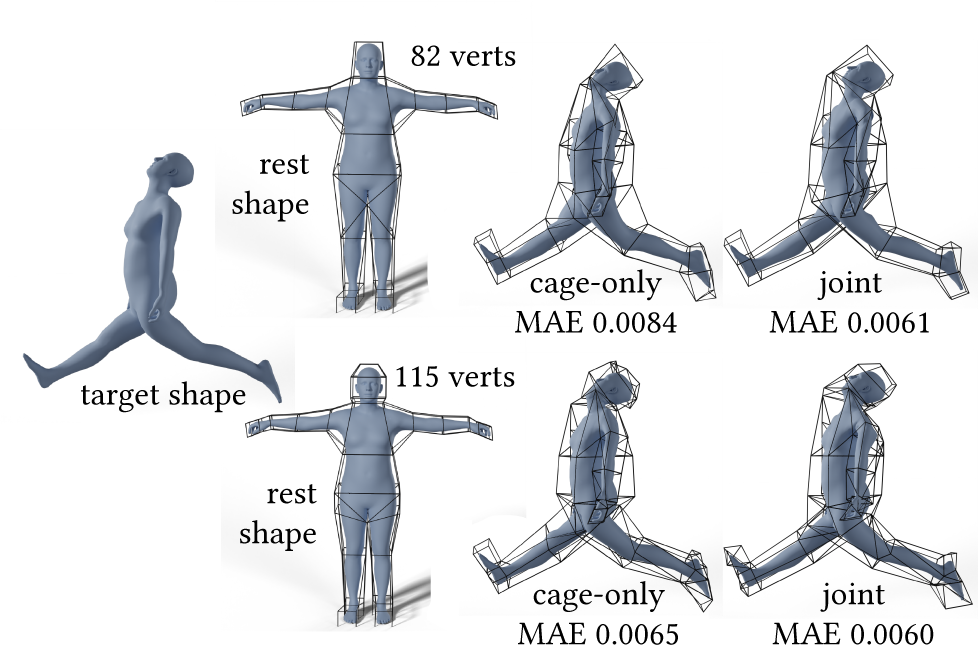}
  \caption{Increasing the cage resolution by $40$\% results in an overall decrease in approximation error. Notably, even when using lower-resolution cage, the joint optimization leads to a $37$\% error reduction.}
  \Description{}\label{fig:invhuman}
\end{figure}

In Figure~\ref{fig:invhuman}, we show the effect of increasing the cage resolution by $40$\% during inverse optimization of a human body mesh \cite{SMPL:2015}. As expected, the additional degrees of freedom introduced by the new cage vertices result in a better fit to the target shape for both the cage-only and the joint cage-and-coordinates optimization. However, as evidenced by the the mean absolute vertex distance, jointly optimizing for the coordinates alongside the cage vertices improves the results significantly even on the lower resolution cage. This implies that the additional degrees of freedom in a cage are not as useful in the presence of coordinates that are better suited for a specific use-case, further validating our deformation-aware approach to designing generalized barycentric coordinates.

In these experiments, we fine-tuned the network for a total of $10000$ steps for the \Armadillo{} mesh, and $15000$ steps for the \Human{} mesh, with an initial learning rate of $5\cdot10^{-4}$. During the first $6000$ steps, we decayed the learning rate by a factor of $0.8$ every $200$ steps. We used a larger learning rate of $5\cdot10^{-3}$ for the cage rotation, scale, and vertex positions.

\section{Discussion and Conclusion}

\paragraph*{Alternative deformation-aware energies.} We have demonstrated how to incorporate the surface as-rigid-as-possible energy into our formulation as a proof-of-concept. 
This work represents the first step towards other deformation-aware barycentric coordinates.
For example, all of the deformation-aware energies we worked with are restricted to the surface of an interior mesh.
Given that the deformation field is volumetric, a reasonable alternative would be to use a volumetric deformation energy, such as the ones enumerated by \citet{abulnaga2022symmetric}.
Since our model is grid-free, it would be necessary to find a way of computing the volumetric deformation gradient required by volumetric deformation energies.

\paragraph*{Alternative inverse problems.} In addition to the inverse deformation problem presented in Figure~\ref{fig:inverse}, our method can help tackle other inverse problems. For example, our experiment requires the existence of dense vertex-to-vertex correspondences between two meshes. This points to a natural direction for future work, namely solving inverse cage-based deformation problems while relying only on sparse correspondences. Pushing this idea further, one might consider combining our inverse deformation framework with an inverse rendering loss, similar to \citet{Peng2021b}, thus completely eliminating the need for $3$D reconstructions or correspondences. As an alternative future direction, one could consider amortizing a physics simulation over multiple frames---or possibly even over an entire dataset---to create coordinates specific to a single class of shapes. 

\paragraph*{Performance.} We see multiple viable avenues for improving the performance of our model. It seems likely that a stochastic finite-difference estimator would reduce the memory and compute cost of computing first-order smoothness energies. Furthermore, our algorithm requires us to find all valid simplices for every interior sample in a brute-force manner. Future work might instead consider applying an acceleration data structure to improve the computational complexity of this step. Lastly, while we already use Tiny CUDA NN \cite{tiny-cuda-nn} in our implementation, we were unable to exploit the fully-fused neural networks offered by the library due to the limitations of our hardware.

\paragraph*{Limitations.} Because of the pruning heuristic from Section~\ref{ss:accel}, our coordinates are limited to a star-shaped neighborhoods of cage vertices. In practice, this means that a slight bend of a straight cage into a concavity can change the coordinates. To address this, one could consider not eliminating virtual simplices that cross the cage boundary, but rather modifying the pruning heuristic to be based on the geodesic distance between simplex vertices instead of edge lengths. However, our method would then no longer be grid-free due to the geodesic distance solver. 

\paragraph*{Conclusion.} Our work departs from existing methods for generalized barycentric coordinates through a fresh theoretical perspective and computational approach.
By relying on the machinery of neural fields, we realize a practical algorithm and demonstrate its usefulness for a number of applications. 
Our work represents a key step toward practical deformation-aware generalized barycentric coordinates.

\begin{acks}
We thank Prof.\ Mirela Ben-Chen for discussion and feedback during the course of this project.

The MIT Geometric Data Processing group acknowledges the generous support of Army Research Office grants W911NF2010168 and W911NF2110293, of Air Force Office of Scientific Research award FA9550-19-1-031, of National Science Foundation grant CHS-1955697, from the CSAIL Systems that Learn program, from the MIT–IBM Watson AI Laboratory, from the Toyota–CSAIL Joint Research Center, from a gift from Adobe Systems, and from a Google Research Scholar award.

The MIT Scene Representation group acknowledges support of the National Science Foundation under grant 2211259, the Singapore DSTA under DST00OECI20300823, the Amazon Science Hub, the Toyota Research Institute, and the MIT-IBM Watson AI Laboratory. 
\end{acks}

\bibliographystyle{ACM-Reference-Format}
\bibliography{bibliography.bib}

\end{document}